\documentclass[a4paper,cite,notitlepage]{article}
\usepackage{latexsym}

\usepackage[T1]{fontenc}
\usepackage[latin1]{inputenc}

\usepackage{graphicx}
\usepackage{color,pst-plot}

\usepackage{amsmath}
\usepackage {amsfonts,amssymb,amstext}

\usepackage{latexsym}

\usepackage{graphicx}
\usepackage{color,pst-plot}

\usepackage{amsmath}
\usepackage {amsfonts,amssymb,amstext}

\newcommand{\lbl}[1]{\label{eq:#1}}
\newcommand{ \rf}[1]{(\ref{eq:#1})}

\newcommand{\be}{\begin{equation}}
\newcommand{\ee}{\end{equation}}
\newcommand{\bea}{\begin{eqnarray}}
\newcommand{\eea}{\end{eqnarray}}
\newcommand{\setl}{\setlength\arraycolsep{2pt}}

\newcommand{\noi}{\noindent}
\newcommand{\nn}{\nonumber}
\newcommand{\ra}{\rightarrow}
\newcommand{\Ra}{\Rightarrow}

\newcommand{\cA}{{\cal A}}

\newcommand{\cF}{{\cal F}}

\newcommand{\cM}{{\cal M}}

\newcommand{\cO}{{\cal O}}

\newcommand{\cR}{{\cal R}}

\newcommand{\Imm}{\mbox{\rm Im}}
\newcommand{\Ree}{\mbox{\rm Re}}

\newcommand{\Li}{\mbox{\rm Li}}

\newcommand{\annd}{\mbox{\rm and}}

\newcommand{\als}{\alpha_{\mbox{\rm {\scriptsize s}}}}

\newcommand{\Meijer}[5]{{\rm G}^{#1}_{#2} \left(#3\left|\begin{matrix} {#4} \\ {#5} \end{matrix}\right.\right) }

\input epsf

\renewcommand{\thesection}{\normalsize \Roman {section}}
\def\theequation{\arabic{section}.\arabic{equation}}

\allowdisplaybreaks[1]

\title{Mellin-Barnes approach to\\
hadronic vacuum polarization and $g_{\mu}-2$}%

\author{J\'{e}r\^{o}me Charles$^1$, David Greynat$^2$ and Eduardo de Rafael$^1$\\
\textit{\small Aix-Marseille Univ, Universit\'e de Toulon, CNRS, CPT, Marseille, France}\\
\textit{\small $^2$ david.greynat@gmail.com}}

\date{\today}
\begin{document}

\maketitle

\begin{abstract}
 It is shown that with a precise determination of  a few derivatives of the hadronic vacuum polarization (HVP) self-energy function  $\Pi(Q^2)$ at $Q^2=0$, from lattice QCD (LQCD) or from a dedicated low-energy experiment, one can  obtain an evaluation of the lowest order HVP contribution to the anomalous magnetic moment of the muon  $a_{\mu}^{\rm HVP}$ with an accuracy comparable to the one reached using the  $e^+ e^-$ annihilation  cross section into hadrons. 
The technique of Mellin-Barnes approximants (MBa) that we propose is illustrated in detail with the example of the two loop vacuum polarization function in QED. We then apply it to the first few moments of the hadronic spectral function obtained from experiment and show that the resulting MBa evaluations of $a_{\mu}^{\rm HVP}$ converge very quickly to the full experimental determination.
 \end{abstract}

\section{\Large Introduction.}
\setcounter{equation}{0}
\def\theequation{\arabic{section}.\arabic{equation}}

\noi
This paper explains and develops the approach recently  described by one of the authors in refs.~\cite{EdeR14,EdeR17a,EdeR17b} to evaluate  the hadronic vacuum polarization (HVP)contribution to the anomalous magnetic moment of the muon $a_{\mu}^{\rm HVP}$. 

Our motivation is threefold:

\begin{enumerate}

\item The persistent discrepancy at the $\sim 4\sigma$ level between the experimental determination of the anomalous magnetic moment of the muon~\cite{BNL}
\be\lbl{eq:exp}
a_{\mu}(\rm E821-BNL)= 116~592~089 (54)_{\rm\tiny stat} (33)_{\rm\tiny syst}\times 10^{-11} [0.54 ppm]\,,
\ee
and the standard model prediction~\cite{TH}	
\be\lbl{eq:sm}
a_{\mu}(\rm SM)= 116~591~805~(42) \times 10^{-11}\,.
\ee

\item
The fact that the standard model contribution which at present has the largest error, is the one coming from the lowest order hadronic vacuum polarization (HVP) 
contribution to $a_{\mu}(\text{SM})$, evaluated from a  combination of experimental results on 
$e^+ e^-$ data~\cite{Davier11,Hagiwara11,Davier16,KNT17}: 
\be\lbl{eq:HVPexps}
a_{\mu}^{\rm HVP}=(6.931\pm 0.034)\times 10^{-8}~{\cite{Davier16}} \quad \annd \quad 
a_{\mu}^{\rm HVP}=(6.933\pm 0.025)\times 10^{-8}~{\cite{KNT17}} \,.
\ee

\item 
The possibility of an alternative evaluation of $a_{\mu}^{\rm HVP}$, either based on QCD first principles with the help of lattice QCD (LQCD) simulations (see e.g. refs.~\cite{Della12}-\cite{lehner17}), or on new dedicated experiments as proposed in ref.~\cite{Venanetal}.
\end{enumerate}

\noi
The standard representation of $a_{\mu}^{\rm HVP}$ used in the experimental determinations is the one in terms of a 
weighted integral of the hadronic spectral function $\frac{1}{\pi}\Imm\Pi(t) $:
\be\lbl{eq:str}
 a_{\mu}^{\rm HVP} = \frac{\alpha}{\pi}\int_{4 m_{\pi}^2}^{\infty}
\frac{dt}{t}\int_{0}^{1}dx\frac{x^2(1-x)}{x^2+\frac{t}{m_{\mu}^2}(1-x)}
\frac{1}{\pi}\Imm\Pi(t)\,.
\ee
Thanks to the optical theorem, the hadronic spectral function is obtained from the total $e^+ e^-$ cross section into hadrons via one photon annihilation ($m_e \ra 0$) 
\be
\sigma(t)_{[e^+ e^- \ra (\gamma)\ra {\rm Hadrons}]}=\frac{4\pi^2 \alpha}{t}\frac{1}{\pi}\Imm\Pi(t)\,.
\ee

We observe that the integrand in Eq.~\rf{eq:str} can be rearranged in a way: 
\be\lbl{eq:eucr}
 a_{\mu}^{\rm HVP}  =  \frac{\alpha}{\pi}\int_{0}^{1}dx\ (1-x)  \int_{4 m_{\pi}^2}^{\infty}
\frac{dt}{t}\ \frac{\frac{x^2}{1-x} m_{\mu}^2}{t+\frac{x^2}{1-x} m_{\mu}^2}\ 
\frac{1}{\pi}\Imm\Pi(t)\,,
\ee
which explicitly displays the dispersion relation between the hadronic spectral function and the { renormalized} hadronic photon self-energy in the euclidean:
\be\lbl{eq:disprel}
-\Pi(Q^2)=\int_{4 m_{\pi}^2}^{\infty}
\frac{dt}{t}\frac{Q^2}{t+Q^2}\frac{1}{\pi}\Imm\Pi(t)\,,\quad{\rm with}\quad 
Q^2 \equiv \frac{x^2}{1-x}m_{\mu}^2 \ge 0 \,,
\ee
and therefore~\cite{LPdeR72,EdeR94}
\be\lbl{eq:eu}
 a_{\mu}^{\rm HVP}  =  -\frac{\alpha}{\pi}\int_{0}^{1}dx\ (1-x)  \ \Pi\left(\frac{x^2}{1-x}m_{\mu}^2\right)\,.
\ee
Trading the Feynman parameter $x$-integration by a $Q^2$-integration results in a slightly more complicated expression 
\be\lbl{eq:lqcd}
a_{\mu}^{\rm HVP}  = 
\frac{\alpha}{\pi}\int_0^\infty \frac{dQ^2}{Q^2} \sqrt{\frac{Q^2}{4 m_{\mu}^2+Q^2}}\left(\frac{\sqrt{4 m_{\mu}^2 +Q^2}-\sqrt{Q^2}}{{\sqrt{4 m_{\mu}^2 +Q^2}+\sqrt{Q^2}}} \right)^2 [-\Pi(Q^2)]\,,
\ee
which is  the one proposed for LQCD evaluations~\cite{Blum03}.
Because of the parametric $x$-dependence in Eq.~\rf{eq:eu},  or the $Q^2$-weight function in the integrand  of Eq.~\rf{eq:lqcd},  the $a_{\mu}^{\rm HVP}$ integral  is dominated by the low-$Q^2$ behaviour of the hadronic self-energy function  $\Pi(Q^2)$. The natural  question which then arises is: 
 {\it What is the best way to help LQCD (see e.g. refs.~\cite{Della12}-\cite{lehner17}), or dedicated experiments~\cite{Venanetal},  to evaluate this integral when only limited information about  $\Pi(Q^2)$ at low $Q^2$ values is available?}
The answer that we propose  follows the way initiated in ref.~\cite{EdeR17a}. It is based on Mellin-Barnes techniques which we shall describe below and which we shall illustrate with several examples. As we shall see, this is a very powerful method compared to other approaches discussed in the literature (see e.g. refs.~\cite{ABCGPT16, BDDJ16, DOMetal17} and references therein). 

The paper has been organized as follows. The next section is an introduction to the QCD properties of the Mellin transform of the HVP spectral function. Section III is dedicated to a few ingredients, which are required to understand and justify the method that we propose. The subsection III.3 is particularly technical since it justifies mathematically the underlying approach and the restriction to the subclass of Marichev-like  Mellin approximants given in Eq.~\rf{eq:marichevend}.  For those who are just interested in the applications, it can be escaped in a first reading.  Section IV illustrates the application of Mellin-Barnes approximants (MBa) to vacuum polarization in QED at the two loop level. Section V tests the advocated technique of MBa with the experimental values of the HVP moments provided to us by the authors of ref.~\cite{KNT17}. These moments, with their errors,  are obtained from the same spectral function which results in the second number quoted in Eq.~\rf{eq:HVPexps}.  
We show how the successive MBa approach the experimental determination of $a_{\mu}^{\rm HVP}$. The conclusions with an outlook on future work are given in Section VI. A few technical details have been included in an Appendix.

\section{\Large The Mellin Transform of the Hadronic Spectral Function.}
\setcounter{equation}{0}
\def\theequation{\arabic{section}.\arabic{equation}}

\noi
In QCD the hadronic spectral function is positive and  goes asymptotically to a constant ($q_i$ denotes the charge, in electric charge units, of an active quark with flavour $i$ ) :
\be\lbl{eq:pqed}
\frac{1}{\pi}\Imm\Pi(t) \underset{{t\ra\infty}}{\thicksim}
\left(\frac{\alpha}{\pi} \right)\left(\sum_i q_i^2\right)\frac{1}{3}N_c \left[1+\cO(\als)\right]\,,
\ee
with perturbative QCD (pQCD) $\als$-corrections known up to four loops.  

The moment integrals
\be\lbl{eq:moments}
\int_{t_0}^\infty\frac{dt}{t}\left(\frac{t_0}{t} \right)^{1+n}\frac{1}{\pi}\Imm\Pi(t)\,,\quad n=0,1,2\cdots \,,
\ee
where throughout the paper $t_0$ denotes the threshold value of the hadronic spectral function:
\be
t_0=4 m_{\pi^{\pm}}^2
\,,
\ee
can be experimentally determined;   and the dispersion relation in Eq.~\rf{eq:disprel} relates them to successive derivatives of the hadronic self-energy  function $\Pi(Q^2)$ at the origin:
\be\lbl{eq:momeucl}
\int\limits_{t_0}^{\infty}\frac{dt}{t}\left(\frac{t_0}{t} \right)^{1+n}
\frac{1}{\pi}\Imm\Pi(t)=
\frac{(-1)^{n+1} }{(n+1)!}( t_0)^{n+1} \left(\frac{\partial^{n+1}}{(\partial Q^2 )^{n+1}}\Pi(Q^2)\right)_{Q^2 =0}\,,\quad n=0,1,2,\cdots\,,
\ee
which are accessible to  LQCD evaluations.
 In fact, as pointed out a long time ago~\cite{BdeR69}, the first moment for $n=0$ provides a rigorous upper bound to the muon anomaly: 
\be
 a_{\mu}^{\rm HVP}
\le  \frac{\alpha}{\pi}\frac{1}{3}\frac{m_{\mu}^2}{t_0}\int_{t_0}^\infty\frac{dt}{t}\ \frac{t_0}{t}\ \frac{1}{\pi}\Imm\Pi(t)
=\left(\frac{\alpha}{\pi}\right)\frac{1}{3}\frac{m_{\mu}^2}{t_0}
\left(-{t_0}\frac{\partial}{\partial Q^2}\Pi(Q^2)\right)_{Q^2 =0}\,.
\ee
Quite generally, the moments in Eq.~\rf{eq:moments} obey constraints which follow from the positivity of the spectral function
and may provide useful tests to LQCD determinations. We discuss these constraints in the Appendix. 

The moment integrals in Eq.~\rf{eq:moments} can be generalized to a function, which is precisely the Mellin transform of the hadronic spectral function $\frac{1}{\pi}\Imm\Pi(t)$ defined as follows~\cite{EdeR14}:
\be\lbl{eq:melspec}
\cM\left[\frac{1}{\pi}\Imm\Pi(t)\right](s)\equiv \cM(s) =\int_{t_0}^\infty\frac{dt}{t}\left(\frac{t}{t_0} \right)^{s-1}\frac{1}{\pi}\Imm\Pi(t)\,,\quad -\infty\le \Ree(s) <1  \,,
\ee
 with the domain of definition extended to the full complex $s$-plane by analytic continuation.
An important property of $\cM$ is that $\cM(-s)$ is a completely monotonic function of $s$, for the real variable $s$ in the interval $]-\infty,1[$. It follows simply from Eq.~\rf{eq:melspec}
 which implies that all the successive derivatives of $M(s)$ satisfy the positivity conditions
\be\lbl{eq:monot}
\cM^{(n)}(s)\ge 0\,\quad{\rm for~all}\quad n\ge 0.
\ee 

As a result,  $\cM(s)$  can have neither poles nor zeros in the negative $\Ree(s)$ axis and has  a perfectly smooth (increasing) shape in this region. This  {\it smoothness property} of $\cM(s)$, which is at the basis of the approximation method that we shall propose, is to be contrasted with the shape  of the spectral function $\frac{1}{\pi}\Imm\Pi(t)$ itself  which, as we know from experiments, has a rather complicated structure.

In QCD, the Mellin transform $\cM(s)$ is singular at $s=1$ with a residue which is fixed by the pQCD asymptotic behaviour of the spectral function in Eq.~\rf{eq:pqed}. The contribution from the $u$, $d$, $s$, $c$, $b$ and $t$ quarks gives
 \be\lbl{eq:QCDMs1}
\cM(s)\underset{{s\ra\ 1}}{\thicksim} \left(\frac{\alpha}{\pi}\right)\left(\frac{4}{9}+\frac{1}{9}+\frac{1}{9}+\frac{4}{9}+\frac{1}{9}+\frac{4}{9}\right)N_c\  \frac{1}{3}\ \frac{1}{1-s}+\cO(\als)\,.
\ee
The spectral function moments are, therefore,  the particular values of the $\cM(s)$ function at $s=0\,, -1\,, -2\,, -N$ with integer $N$.

As discussed in refs.~\cite{EdeR14,EdeR17a} there exists a representation of $\Pi(Q^2)$, and hence of the anomaly $ a_{\mu}^{\rm HVP}$, in terms of the Mellin transform $\cM(s)$. This follows from inserting the Mellin-Barnes identity~\footnote{For the benefit of the reader who may be unfamiliar with Mellin-Barnes integrals we give a proof of this identity in the Appendix.}
\be\lbl{eq:MBaid}
\frac{1}{1+\frac{Q^2}{t}}=\frac{1}{2\pi i}\int\limits_{c_s-i\infty}^{c_s+i\infty}ds\ \left(\frac{Q^2}{t}\right)^{-s}\ \Gamma(s)\Gamma(1-s)
\ee
in the dispersion relation in Eq.~\rf{eq:disprel},   which results in the representation
\be\lbl{eq:MBaPI}
\Pi(Q^2) = -\frac{Q^2}{t_0}\ \frac{1}{2\pi i}\int\limits_{c_s-i\infty}^{c_s+i\infty}ds\ \left(\frac{Q^2}{t_0} \right)^{-s} \Gamma(s)\Gamma(1-s)\ \cM(s)\,,\quad  c_s \equiv \Ree(s) \in ]0,1[  \,;
\ee
and the corresponding integral representation for the Adler function 
\be\lbl{eq:adler}
\cA(Q^2)\equiv -Q^2 \frac{\partial \Pi(Q^2)}{\partial Q^2}
=\frac{1}{2\pi i}\int\limits_{c_s-i\infty}^{c_s+i\infty}ds\ \left(\frac{Q^2}{t_0} \right)^{1-s} \Gamma(s)\Gamma(2-s)\ \cM(s)\,,\quad  c_s \equiv \Ree(s) \in ]0,1[  \,.
\ee

\noi
Setting $Q^2 =\frac{x^2}{1-x}m_{\mu}^2$ in the representation of $\Pi(Q^2)$ in Eq.~\rf{eq:MBaPI} and inserting it in the r.h.s. of Eq.~\rf{eq:eu} we have

{\setl
\bea
 a_{\mu}^{\rm HVP} & = &  -\frac{\alpha}{\pi}\int_{0}^{1}dx\ (1-x)  \ \Pi\left(\frac{x^2}{1-x}m_{\mu}^2\right)\\
 & = & \frac{\alpha}{\pi}\int_{0}^{1}dx\ (1-x)
\ \frac{1}{2\pi i}\int\limits_{c_s-i\infty}^{c_s+i\infty}ds\ \left(\frac{\frac{x^2}{1-x}m_{\mu}^2 }{t_0} \right)^{1-s} \Gamma(s)\Gamma(1-s)\ \cM(s)\,.
\eea}

\noi
The integral over the $x$-parameter can now be made analytically, leading to the expression~\cite{EdeR14}
\be\lbl{eq:MBamu}
a_{\mu}^{\rm HVP} =   \left(\frac{\alpha}{\pi}\right) \frac{m_{\mu}^2}{t_0}\frac{1}{2\pi i}\int\limits_{c_s -i\infty}^{c_s +i\infty}ds\left(\frac{m_{\mu}^2}{t_0} \right)^{-s} \cF(s)\ \cM(s)\,,\quad  c_s \equiv \Ree(s) \in ]0,1[\,, 
\ee
where $\cF(s)$ is a product of three Gamma functions: 
\be
\cF(s)= -\Gamma(3-2s)\ \Gamma(-3+s)\ \Gamma(1+s)\,,
\ee
and the hadronic dynamics is thus entirely factorized in the Mellin transform $\cM(s)$.

\begin{figure}[!ht]
\begin{center}
\hspace*{-1cm}\includegraphics[width=0.50\textwidth]{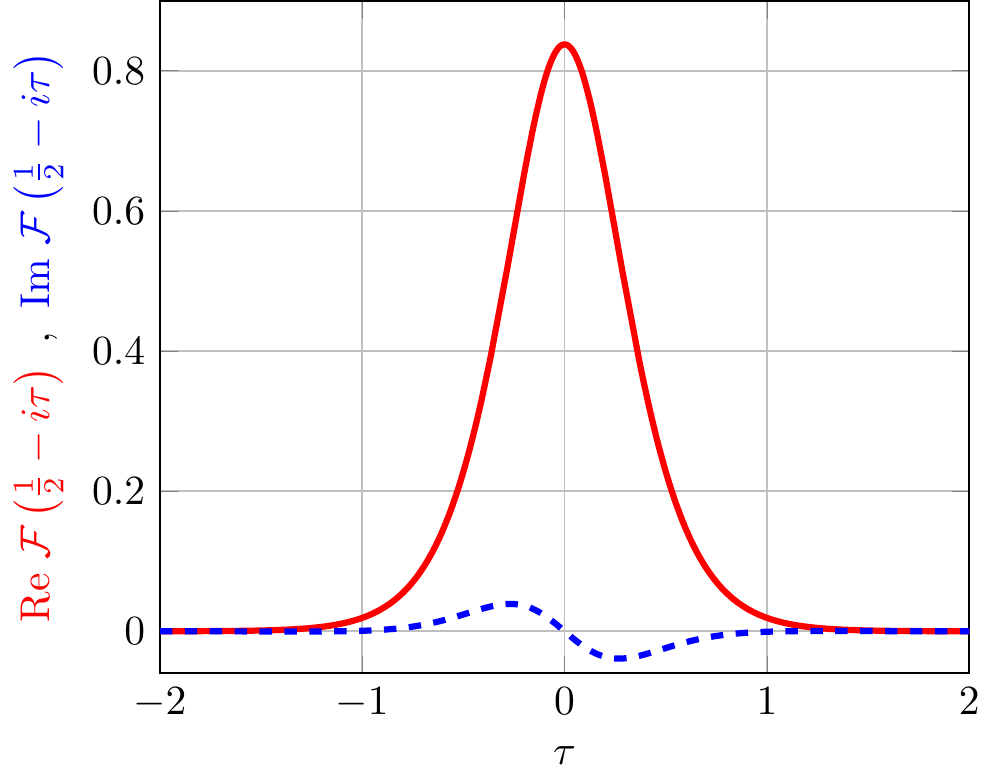} 
\bf\caption{\lbl{fig:Ftau}} 
\vspace*{0.25cm}
{\it Shape of the function $\cF\left(\frac{1}{2}-i\tau\right)$ in Eq.~\rf{eq:MBamu} versus $\tau$.\\ The red curve is the real part of the function, the blue dashed curve its imaginary part.}
\end{center}
\end{figure}

\noi
The weight function $\cF(s)$ in Eq.~\rf{eq:MBamu} is universal and has a shape which, for  $s$ within the {\it fundamental strip}~\cite{FGD95}: $c_s \equiv \Ree(s) \in ]0,1[$ and the choice $s=\frac{1}{2}-i\tau$, is shown in Fig.~\rf{fig:Ftau} as a function of $\tau$. Notice that the real part of this function (the red curve) is symmetric under $\tau\ra -\tau$ while its imaginary part is antisymmetric. Both the real and imaginary parts fall  very fast as $\tau$ increases. 
With the change of variable 
\be
s\ra \frac{1}{2}-i\tau\,,
\ee
the integral in Eq.~\rf{eq:MBamu} becomes then a Fourier transform:
\be\lbl{eq:MBamuF}
a_{\mu}^{\rm HVP} =   \left(\frac{\alpha}{\pi}\right)\sqrt{\frac{m_{\mu}^2}{t_0}}\frac{1}{2\pi }\int\limits_{-\infty}^{+\infty}d\tau \  e^{-i\tau \log\frac{t_0}{m_{\mu}^2}}\  \cF\left(\frac{1}{2}-i\tau\right)\ \cM\left(\frac{1}{2}-i\tau \right)\,. 
\ee
Because of the shape of the $\cF\left(\frac{1}{2}-i\tau\right)$ function and the growth restrictions on $\cM\left(\frac{1}{2}-i\tau \right)$ for large $\tau$, which are fixed by the fact that  $\Pi(Q^2)$ obeys a dispersion relation in QCD, this Fourier integral is fully dominated by the behaviour of the integrand   in a very restricted $\tau$-interval, $-T\le \tau \le +T$ with $T$ of order one.

\section{\Large Some Technical Ingredients.}
\setcounter{equation}{0}
\def\theequation{\arabic{section}.\arabic{equation}}

\noi
We shall next recall a few technical ingredients which in the literature go under the name of: Ramanujan Master Theorem, Marichev class of Mellin transforms, Generalized Hypergeometric Functions and Meijer's G-Functions. They are necessary to implement and justify  the MBa framework that we propose.

\subsection{\large The so called Ramanujan's Master Theorem.}

\noi
Consider a function $F(x)$ which admits a power series expansion
\be
F(x)\underset{{x\ra 0}}{\thicksim} \lambda(0)-\lambda(-1)x+\lambda(-2)x^2-\lambda(-3)x^3 +\cdots\,.
\ee
Ramanujan's theorem refers then to the formal identity~\cite{Berndt85}
\be
\int_0^\infty\ dx\ x^{s-1} \left\{ \lambda(0)-\lambda(-1)x+\lambda(-2)x^2-\lambda(-3)x^3 +\cdots\right\}=\Gamma(s)\Gamma(1-s)\lambda(s)\,,
\ee
and implies that the Mellin transform of $F(x)$ is given by
\be
\int_0^\infty dx x^{s-1} F(x) = \Gamma(s)\Gamma(1-s)\lambda(s)\,.
\ee
The function $\lambda(s)$, extended over the full complex $s$-plane, can thus be  simply obtained from the discrete $n$-functional dependence of the $\lambda(-n)$ coefficients  of the Taylor expansion of $F(x)$ by the formal replacement $n\ra -s$. The proof of this beautiful theorem was provided by Hardy~\cite{Hardy78} and it is based on Cauchy's residue theorem as well as on the Mellin-Barnes representation. The basic assumption in Hardy's proof is a growth restriction on $|\lambda(s)|$ which assures that the series $\lambda(0)-\lambda(-1)x+\lambda(-2)x^2-\lambda(-3)x^3 +\cdots$ has some radius of convergence. In our case $F(x)$ will be the hadronic photon self-energy function $\Pi(Q^2)$, with $x\equiv\frac{Q^2}{t_0}$, and Hardy's growth restriction is equivalent to the one required to write a dispersion relation for $\Pi(Q^2)$. 

At small $Q^2$ values, the hadronic photon self-energy function $\Pi(Q^2)$ in QCD has indeed a power series expansion:
\be
-\frac{t_0}{Q^2}\Pi(Q^2)\underset{{Q^2\ra 0}}{\thicksim}\ 
\cM(0)-\frac{Q^2}{t_0}\cM(-1)+\left(\frac{Q^2}{t_0}\right)^2 \cM(-2)-\left(\frac{Q^2}{t_0}\right)^3 \cM(-3)+\cdots \,,
\ee
and the coefficients $\cM(0)$, $\cM(-n)$, $n=1,2,3,\dots$ are precisely the moments of the spectral function  defined in Eq.~\rf{eq:momeucl}. Ramanujan's theorem implies then  that
\be
\int_0^\infty d\left(\frac{Q^2}{t_0}\right)\left(\frac{Q^2}{t_0}\right)^{s-1}\left\{\cM(0)-\frac{Q^2}{t_0}\cM(-1)+\left(\frac{Q^2}{t_0}\right)^2 \cM(-2)+\cdots\right\}
 = \Gamma(s)\Gamma(1-s)\ \cM(s)\,,
\ee
which allows, in principle,  to reconstruct the Mellin transform $\cM(s)$ in the full complex $s$-plane  from just the knowledge of  the discrete moments $\cM(-n)$, $n=0,1,2,3,\cdots$. Given $N$ moments $\cM(-n)$, $n=0,1,2,3,\cdots N-1$, the method of Mellin-Barnes approximants (MBa) that we propose constructs successive $\cM_{N}(s)$ functions which exactly reproduce the values of the first $N$-moments and  approximate better and better the full $\cM(s)$. When inserted in the integrand of the r.h.s. of Eq.~\rf{eq:MBamuF} they result in a set of successive $a_{\mu}^{\rm HVP}(N)$ approximations to the full $a_{\mu}^{\rm HVP}$. A simple example of this procedure was discussed in ref.~\cite{EdeR17a} in the case of vacuum polarization in QED at the one loop level where, in that case, the corresponding Mellin transform is exactly reproduced from its knowledge at just three $s$ values: e.g. $s=1,0,$ and $-1$.

\subsection{\large Marichev's Class of Mellin Transforms.}

\noi
The class in question is the one  defined by {\it standard  products} of gamma functions of the type
\be\lbl{eq:marichev}
\cM(s)=C\ \displaystyle\prod_{i,j,k,l}\frac{\Gamma(a_{i}-s)\Gamma(c_{j}+s)}{\Gamma(b_{k}-s)\Gamma(d_{l}+s)}\,,
\ee
with constants $C$, $a_i$, $b_k$, $c_j$ and $d_l$ and where the Mellin variable $s$ only appears with a $\pm$ coefficient. 
The interesting thing about this class of functions is that all the Generalized Hypergeometric Functions have Mellin transforms of this type~\cite{Marichev83}. As a result, many functions have a representation in terms of  Mellin-Barnes integrals involving linear combinations of standard products of the  Marichev type in Eq.~\rf{eq:marichev}.~\footnote{For a helpful tutorial see e.g. ref.~\cite{Fikioris06} and references therein.}

 In our case, the monotonicity property in Eq.~\rf{eq:monot} of the QCD Mellin transform  implies  precise restrictions on the subclass of Marichev-like functions that one must consider when trying to implement successive approximations.
In that respect we have been particularly helped by some relatively recent mathematical literature~\cite{PK01,PTZ94,PASS96}. The authors of these references have studied the general conditions for the convergence of a very general class of Mellin-Barnes integrals, which include those of the Marichev class, and their results can be summarized as follows.

Consider the rather general type of Mellin-Barnes integral
\begin{equation}\lbl{eq:generalint}
I(z) =\frac{1}{2\pi i} \int\limits_{c-i\infty}^{c+i\infty} ds \,
z^{-s}\frac{\prod_{j=1}^m\Gamma(A_j s + B_j )}{\prod_{k=1}^n\Gamma(C_k s + D_k )}\,.
\end{equation}
In our case this will apply to the Mellin-Barnes integral in  Eq.~\rf{eq:MBaPI} where
\be
 z\equiv\frac{Q^2}{t_0}\quad\annd\quad I(z)\equiv -\frac{t_0}{Q^2}\Pi(Q^2)\,,
\ee
as well as to the Mellin-Barnes integral in Eq.~\rf{eq:MBamu} where 
\be
 z\equiv\frac{m_{\mu}^2}{t_0}\quad\annd\quad I(z)\equiv a_{\mu}^{\rm HVP}(z)\,.
\ee
Quite generally, the authors of refs.~\cite{PK01,PTZ94} have studied the properties of the mapping which integrals like those in Eq.~\rf{eq:generalint} establish between the Mellin $s$-plane and the $z$-plane. This  is illustrated in Fig.~\rf{fig:mapping} where the crosses denote the positions of the poles in the integrand of Eq.~\rf{eq:generalint}: in blue the poles at the left of the fundamental strip (represented by the  green strip in the figure) and in red at the r.h.s. of the fundamental strip. In the $z$-plane we show the disc $\vert z\vert \le R$ in blue, with $R$ the radius of convergence, and the cut starting at $\Ree (z)\ge R$~\footnote{For the sake of simplicity in drawing  the figure, we assume that the disc of convergence is centered at $z=0$ and that the cut starts at $\Ree (z)\ge R$.}. The  converse mapping theorem of ref.~\cite{FGD95} relates in a precise way the singularities in the complex $s$-plane 
of the integrand in Eq.~\rf{eq:generalint}  to the  asymptotic expansions of  $I(z)$ for $z$ large (the red mapping in Fig.~\rf{fig:mapping}) and for $z$ small (the blue mapping in Fig.~\rf{fig:mapping}).
\begin{figure}[!ht]
\begin{center}
\hspace*{-1cm}\includegraphics[width=0.70\textwidth]{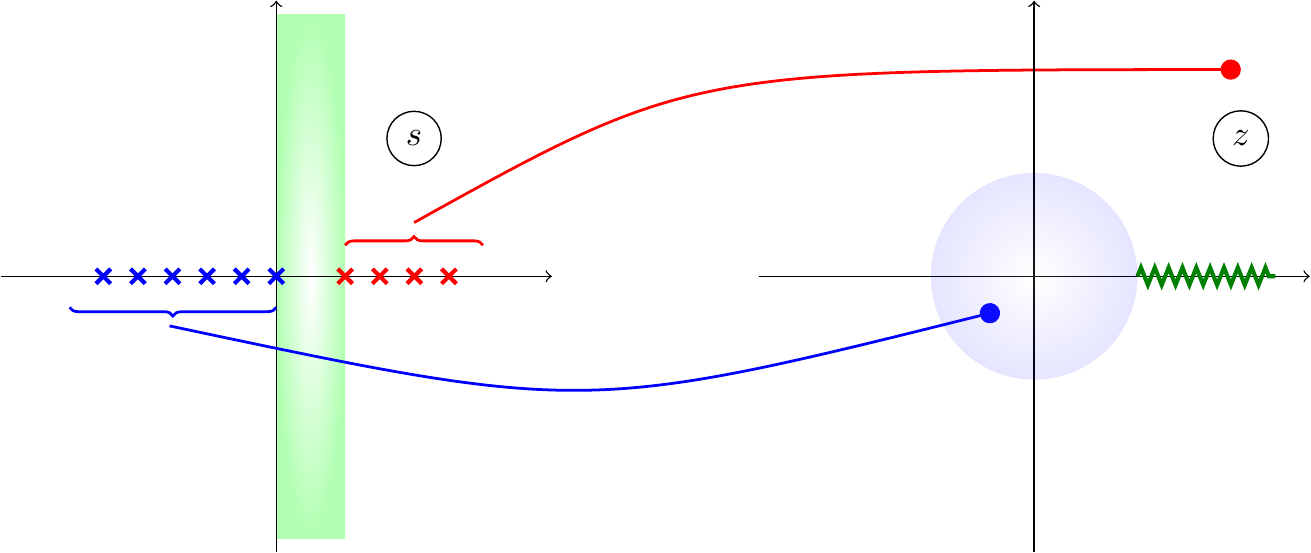}
\bf\caption{\lbl{fig:mapping}} 
\vspace*{0.25cm}
{\it Mapping of the Mellin $s$-Plane to the $z$-plane.} 
\end{center}
\end{figure}
\noi
Following refs.~\cite{PK01,PTZ94,PASS96} we are instructed to consider the two quantities:
\be
 \Delta  \doteq  \sum_{j=1}^m A_j-\sum_{k=1}^n C_k\quad\annd\quad  \alpha  \doteq  \sum_{j=1}^m|A_j|-\sum_{k=1}^n|C_k|\,.
\ee

\noi
Then, the region where the integral $I(z)$ converges is $|\arg z|<\frac{\pi}{2}\alpha$~(see e.g. \cite{PK01}), and there are three cases to be considered~\cite{PTZ94,PASS96}:

\begin{itemize}
\item If $\Delta>0$, closing the integration contour to the left leads to a series representation of the integral $I(z)$ which converges for any value of $z$, but closing the contour to the right gives a divergent asymptotic expansion.

\item If $\Delta<0$, closing the contour to the right leads to a series representation of $I(z)$ which converges for any value of $z$, but closing the contour to the left gives a divergent asymptotic expansion.

\item If $\Delta=0$, closing the contour to the left and to the right gives two convergent series, the first series obtained by closing to the left converges within  a disk $|z|<R$ whereas the other one converges outside this disk. Moreover, if $\alpha>0$, the two series are the analytic continuation of each other.
\end{itemize}

\noi
These three cases are illustrated in Fig.~\rf{fig:threecases}.

\begin{figure}[!ht]
\begin{center}
{\includegraphics[width=0.90\textwidth]{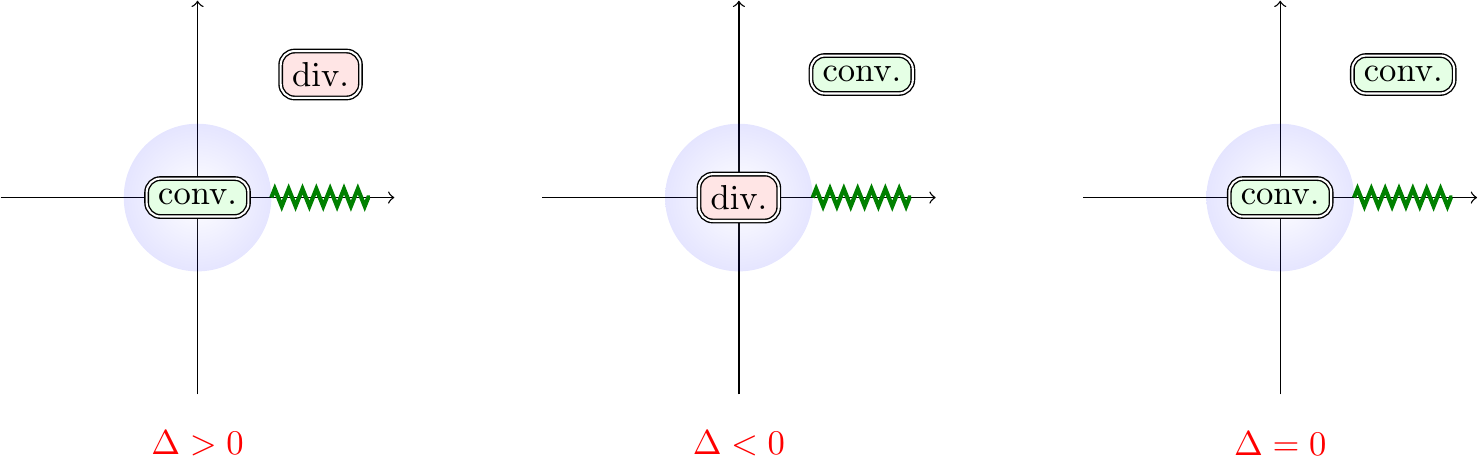}}
\bf\caption{\lbl{fig:threecases}} 
\vspace*{0.25cm}
\end{center}
{\it Behaviour of the series expansions of $I(z)$ depending on the sign of $\Delta$ for $|z|<R$ (the blue region) and $|z|>R$. The label \textit{div.} denotes  the regions where the asymptotic expansion is divergent or does not exist. The cut is represented by the green zigzag line.}
\end{figure}

We are now in the position of fixing the class of successive Mellin approximants $\cM_{N}(s)$ that we { should} use to ensure that they converge in the same way as the full QCD Mellin transform $\cM(s)$ does. Associated to each $\cM_{N}(s)$ approximant there will be a corresponding $\Pi_{N}(Q^2)$ approximant to $\Pi(Q^2)$ { (via Eq.~\rf{eq:MBaPI})} and, therefore, a corresponding $a_{\mu}^{\rm HVP}(N)$ approximant  to $a_{\mu}^{\rm HVP}$ { (via Eq.~\rf{eq:MBamuF})}. 
The input will be that we know the values of the first few  moments 
\be\lbl{eq:input}
\cM(0)\,,\quad\cM(-1)\,,\quad\cM(-2)\,,\cdots\,, \quad\cM(-N+1)\,,
\ee
including their errors and their correlation matrix, 
either from a LQCD determination or from a dedicated experiment. 
Given this input, we shall { then} restrict the successive Marichev-like Mellin approximants in Eq.~\rf{eq:marichev} to those satisfying the following criteria:

\begin{enumerate}

	\item
The fundamental strip of each Mellin approximant $\cM_{N}(s)$ must be the same as the one of the full Mellin transform $\cM(s)$, so that the insertion of  $\cM_{N}(s)$ in the r.h.s. of Eq.~\rf{eq:MBamu} does not change the convergence region $c_s \equiv \Ree(s) \in ]0,1[$ of the exact Mellin transform. 
	
	In practice, due to the fact that the sequence of poles from $\Gamma(a_i -s)$ is at $s=a_i +n$ and the one from $\Gamma(c_j +s)$ at $s=-c_j-n$ with $n\in{\bf N}$ implies the restrictions:
	\be
	\Ree~a_i \ge 1\quad \annd\quad \Ree~c_j \ge 0\,.
	\ee

\item	
 The Mellin approximant $\cM_{N}(s)$ should not generate poles nor zeros in the region $-\infty<\Ree(s)<1$, where $\cM(s)$ is known to be monotonously increasing.  Since $\Ree~c_j \ge 0$,  no poles for  $\Ree(s)<1$ 
 implies the absence of  factors $\Gamma(c_j +s)$ or $j_{\rm max}=0$. No zeros for $\cM_{N}(s)$ in the region $-\infty<\Ree(s)<1$ implies
\be
\Ree~b_k \ge 1. 
\ee

\item
 { We also want the corresponding $\Pi_{N}(Q^2)$-function (see Eq.~\rf{eq:MBaPIN} below) to the Mellin approximant $\cM_{N}(s)$} to converge for $z\equiv \frac{Q^2}{t_0}$ both for $|z|<1$ and $|z|>1$ which, according to the convergence conditions discussed above, requires that 
\be\lbl{eq:delta}
\Delta =(1-1-i_\mathrm{max})-(-k_\mathrm{max}+l_\mathrm{max})=k_\mathrm{max}-i_\mathrm{max}-l_\mathrm{max}=0\,.
\ee

\item
Finally, we want the two series  generated by  the $\Pi_{N}(Q^2)$ approximant for $|z|<1$ and $|z|>1$ to be the analytic continuation of each other which implies
\be
\alpha=(2+i_\mathrm{max}) -(k_\mathrm{max}+l_\mathrm{max})>0 \,.
\ee
This, combined with Eq.~\rf{eq:delta}, implies $l_\mathrm{max}<1$ and hence the absence of $\Gamma(d_l +s)$ factors in the denominator of Eq.~\rf{eq:marichev}.
  
\end{enumerate}

\noi
From the above considerations we conclude that, in the case of HVP in QCD,  the only  Mellin approximants of the  Marichev  class that one must consider are those restricted to the subclass:
\be\lbl{eq:marichevend}
\cM_{N}(s)=C_{N}\ \displaystyle\prod_{k=1}^{N}\frac{\Gamma(a_{k}-s)}{\Gamma(b_{k}-s)}\,,
\ee
with $C_N >0$ and both
\be\lbl{eq:akbk}
\Ree~a_k \ge 1\quad\annd\quad\Ree~b_k \ge 1\,.
\ee
Furthermore, the monotonicity property of the QCD Mellin transform requires that (see e.g. ref.~\cite{Alzer97}) 
\be\lbl{eq:bkmak}
\lambda_{N}\doteq\sum_{k=1}^{N}\left(b_{k}-a_{k}\right)\ge 0\,,
\ee
which implies the asymptotic behaviour
\be
\cM_{N}(s)\underset{{s\ra -\infty}}{\thicksim}C_{N}(-s)^{-\lambda_{N}}\,,
\ee
and assures the positivity of $\cM_{N}(s)$ for $\Ree(s)\in ]-\infty,1[$.

When considering a linear superposition of functions of the subclass in Eq.~\rf{eq:marichevend}:
\be
\cM_{N_{1}}+\cM_{N_{2}}+\cdots\,,
\ee
each term must satisfy the restrictions in Eqs.~\rf{eq:akbk} and \rf{eq:bkmak} with real constants $C_{N_{1}}\,,C_{N_{2}}\,,\cdots$ such that
\be
C_{N_{1}}+C_{N_{2}}+\cdots\ge 0\,.
\ee

Besides the matching to the input moments in Eq.~\rf{eq:input}, all the MBa that we shall use will be constrained to satisfy the leading pQCD short-distance behaviour~\footnote{It is possible to incorporate $\alpha_s$ corrections as well. They don't change, however,  the residue of the pole at $s=1$.} 
\be\lbl{eq:pQCD1}
\cM^{\rm QCD}(s)\underset{{s\ra 1}}{\thicksim} 	\frac{\alpha}{\pi}\left(\sum_i q_i^2\right)\frac{1}{3}N_c~\frac{1}{1-s}\,.
\ee
Given a MBa $\cM_{N}(s)$, the corresponding $\Pi_{N}(Q^2)$ approximant to $\Pi(Q^2)$ is then
\be\lbl{eq:MBaPIN}
\Pi_{N}(Q^2) = -\frac{Q^2}{t_0}\ \frac{1}{2\pi i}\int\limits_{c_s-i\infty}^{c_s+i\infty}ds\ \left(\frac{Q^2}{t_0} \right)^{-s} \Gamma(s)\Gamma(1-s)\ \cM_{N}(s)\,,\quad  c_s \equiv \Ree(s) \in ]0,1[  \,,
\ee
and the $a_{\mu}^{\rm HVP}(N)$ approximant to $a_{\mu}^{\rm HVP}$ is given by the integral in Eq.~\rf{eq:MBamuF} with the corresponding
$\cM_{N}\left(\frac{1}{2}-i\tau \right)$ inserted in the r.h.s. of the integrand.
Notice that the factor $\cF(s)$ does not modify the convergence criteria discussed above for $a_{\mu}^{\rm HVP}(N)$ because $\cF(s)$  has $\Delta=0$ and $\alpha=4$.

\subsection{\large The $\Pi_{N}(Q^2)$  are Generalized Hypergeometric Functions.\\ 
The $\Imm\Pi_{N}(t)$ are Meijer's G-Functions~\protect\footnote{These special functions are built-in in several computer languages. Our definition is consistent with \textit{Mathematica} software that we have used to perform the numerical analyses.}.}

\noi
The Generalized Hypergeometric Function~\cite{Erdely53} 
\be
{_P}{F}_{Q}[a_1 ,a_2  ,\dots a_P ; b_1 ,b_2  ,\dots b_Q ; z] ~\equiv~ _{P}{F}_{Q}\left(\left. \begin{array}{cccc} a_1 & a_2 & \dots & a_P \\ b_1 & b_2 & \dots & b_Q \end{array}\right\vert {z}\right)\,,
\ee
is defined, for $\vert z\vert <1$, by the series

{\setl
\bea\lbl{eq:hgseries}
\lefteqn{ \hspace*{-2cm} 1+\frac{a_1 a_2 \dots a_P}{b_1 b_2 \dots b_Q}\frac{z}{1!}
+\frac{a_1 (a_1 +1) a_2 (a_2 +1) \dots a_P (a_P +1)}{b_1 (b_1 +1) b_2 (b_2 +1) \dots b_Q (b_Q +1)}\frac{z^2}{2!}+\cdots}\nn \\
& & 
\equiv \sum_{n=0}^\infty \frac{(a_1 )_{n}(a_2 )_{n}\dots (a_P )_{n}}{(b_1 )_{n}(b_2 )_{n}\dots (b_Q )_{n}}\frac{z^n}{n!}\,,
\eea}

\noi
where in the second line we use the Pochhammer symbol
\be
(a)_n \equiv \frac{\Gamma(a+n)}{\Gamma(a)}=a(a+1)(a+2)\cdots (a+n-1)\,,
\ee
with in particular,
\be
(a)_0 =1\,,\quad\annd\quad (1)_n =n!\,.
\ee
This series has $P$ numerator parameters, $Q$ denominator parameters and one variable $z$. Any of these parameters are real or complex, but the $b$ parameters must not be negative integers. The case where $P=2$ and $Q=1$ corresponds to the so called Gauss Hypergeometric Function.  The sum of this type of  series, when it exists, defines a Generalized Hypergeometric Function (GH-Function).

The reason why we are interested in GH-Functions is that,
inserting the general expression in Eq.~\rf{eq:marichevend} for the $\cM_{N}(s)$ approximant in the integrand of the  r.h.s. in Eq.~\rf{eq:MBaPIN}, and then doing the  Mellin-Barnes integral over the $s$-variable, results in a specific GH-Function of the type:
\begin{equation}\lbl{eq:GHF}
\Pi_{N}(Q^2)=-\frac{Q^2}{t_0}\ C_N\  \displaystyle\prod_{k=1}^{N}\frac{\Gamma(a_{k})}{\Gamma(b_{k})}\ _{1+N}{F}_{N}\left(\left. \begin{array}{ccccc} 1  & a_1 \dots & a_N \\ ~  & b_1  \dots & b_N \end{array}\right\vert {-\frac{Q^2}{t_0}}\right)\,,
\end{equation}
which is given by the series in Eq.~\rf{eq:hgseries} for $\vert \frac{Q^2}{t_0}\vert <1$, with its analytic continuation defined by the underlying Mellin-Barnes integral, Eq.~\rf{eq:MBaPIN} in this case. The corresponding Adler function is also a GH-Function:
\begin{equation}\lbl{eq:AdlerGHF}
\cA_{N} (Q^2)\equiv -Q^2 \frac{\partial \Pi_{N}(Q^2)}{\partial (Q^2)}=\frac{Q^2}{t_0}\ C_N\  \displaystyle\prod_{k=1}^{N}\frac{\Gamma(a_{k})}{\Gamma(b_{k})}\ _{1+N}{F}_{N}\left(\left. \begin{array}{ccccc} 2  & a_1 \dots & a_N \\ ~  & b_1  \dots & b_N \end{array}\right\vert {-\frac{Q^2}{t_0}}\right)\,.
\end{equation}

The reason why we are interested in Meijer's G-Functions is that the inverse Mellin transform of $\cM_{N}(s)$ corresponding to Eq.~\rf{eq:melspec}, i.e. the Mellin Barnes integrals
\be
\frac{t_0}{t}\frac{1}{\pi}\Imm\Pi_{N}(t)=\frac{1}{2\pi i}\int\limits_{c-i\infty}^{c+i\infty}ds \left(\frac{t}{t_0} \right)^{-s} \cM_{N}(s)\,,\quad c_s \equiv \Ree(s) \in ]-\infty,1[ \,, 
\ee
for arbitrary $N$ and $t\ge t_0$ 
are a particular class of  Meijer's G-Functions.
Indeed, in full generality,  Meijer's G-Functions are defined by a complex $L$-path integral~(see e.g. {\it The Meijer G-Function $\Meijer{m,n}{p,q}{z}{ \,\boldsymbol{a} }{ \,\boldsymbol{b}}$}, in sect.~8.2 of ref.~\cite{PMB90}, pp.~617-626):

{\setl
\bea
\lefteqn{\Meijer{m,n}{p,q}{z}{1-a_1,\ldots, 1-a_n\,;a_{n+1},\ldots,a_p}{b_1,\ldots,b_m\,;\,1-b_{m+1},\ldots,1-b_q}=} \nn \\  & & 
\frac{1}{2\pi i} \int_{L} ds\ z^{-s} \ \frac{\Gamma(b_1+s)\cdots\Gamma(b_m+s)\cdot\Gamma(a_1-s)\cdots\Gamma(a_n-s)}{\Gamma(a_{n+1}+s)\cdots\Gamma(a_p+s)\cdot\Gamma(b_{m+1}-s)\cdots\Gamma(b_q-s)}\,,
\eea}

\noi
and have the property that
\begin{equation}
\Meijer{0,n}{p,q}{z}{ \,\boldsymbol{a} }{ \,\boldsymbol{b}} = 0  \; \; \text{for} \;\; |z|<1\;.
\end{equation}
For the class of Marichev-like $\cM_{N}(s)$ functions in Eq.~\rf{eq:marichevend} this results in a set of {\it equivalent} spectral functions:  
\begin{equation}
\frac{1}{\pi} \Imm\Pi_{N}(t) = \frac{t}{t_0}\; C_N \;\Meijer{0,N}{0,N}{\frac{t}{t_0}}{1-a_1,\ldots,1-a_N \,; \,\relbar\!\relbar}{\relbar\!\relbar\, ;\, 1-b_1,\cdots,1-b_N}\;.
\end{equation}
These successive {\it equivalent} spectral functions, alike the physical spectral function,  are only defined for $t\ge t_0$ but they are not expected to reproduce, {\it locally}, the detailed physical shape unless the level of approximation reaches the exact solution (as it is the case in the QED example at the one loop level discussed in ref.~\cite{EdeR17a}). However, when inserted in a dispersion relation integral, they reproduce the predicted smooth behaviour of the successive self-energy functions $\Pi_{N}(Q^2)$ and  Adler $\cA_{N}(Q^2)$ functions. It is in this sense that we call them {\it equivalent}.

The explicit form of these general expressions for the first $N=1$ and $N=2$ cases are as follows:

\begin{itemize}

\item N=1
	
This corresponds to the case where we only know the first moment $\cM(0)$. Then
\be
\cM_1 (s)=C_1 \frac{\Gamma(a_1-s)}{\Gamma(b_1 -s)}\,,\quad{\rm with}\quad C_1= \frac{\alpha}{\pi}\frac{5}{3}\frac{N_c}{3}\Gamma(b_1 -1)\quad\annd\quad a_1=1
\ee
to ensure the pQCD pole behaviour at $s=1$. The only free parameter $b_1$  is then fixed by the matching condition $\cM_1 (0)=\cM(0)$ and  one finds
\be
\Pi_1 (Q^2)=-\frac{Q^2}{t_0}\ C_1 \frac{1}{\Gamma(b_1)}
\ _{2}{F}_{1}\left(\left. \begin{array}{cc} 1 & a_1 \\ ~ & b_1\end{array}\right\vert {-\frac{Q^2}{t_0}}\right)\,,
\ee
and the corresponding Adler function [see Eq.~\rf{eq:adler}] is 
\be
\cA_1 (Q^2)=-Q^2 \frac{\partial \Pi_1(Q^2)}{\partial (Q^2)}=
\frac{Q^2}{t_0}\ C_1 \frac{1}{\Gamma(b_1)}
\ _{2}{F}_{1}\left(\left. \begin{array}{cc} 2 & a_1 \\ ~ & b_1\end{array}\right\vert {-\frac{Q^2}{t_0}}\right)\,.
\ee
In this simple case the {\it equivalent} spectral function is

{\setl
\bea
\frac{1}{\pi} \Imm\Pi_{1}(t) & = &  \frac{t}{t_0}\; C_1 \;\Meijer{0,1}{0,1}{\frac{t}{t_0}}{1-a_1 \,; \,\relbar\!\relbar}{\relbar\!\relbar\, ;\, 1-b_1}\\
 & = & \frac{\alpha}{\pi}\frac{5}{3}\left(\frac{t_0}{t} \right)^{a_1 -1}\left(1-\frac{t_0}{t} \right)^{b_1-2}\,.
\eea}

\item N=2

This corresponds to the case where we know the first two moments $\cM(0)$ and $\cM(-1)$. Then
\be
\cM_2 (s)=C_2\frac{\Gamma(1-s)}{\Gamma(2 -s)}\frac{\Gamma(a_2 -s)}{\Gamma(b_2-s)}\quad{\rm with}\quad C_2= \frac{\alpha}{\pi}\frac{5}{3}\frac{N_c}{3}\frac{\Gamma(b_2 -1)}{\Gamma(a_2 -1)}\,,
\ee
and the parameters $a_2$ and $b_2$ fixed by the two matching conditions
\be
\cM_2 (0)=\cM(0)\quad\annd\quad\cM_2 (-1)=\cM(-1)\,.
\ee 
Then
\be
\Pi_2 (Q^2)=-\frac{Q^2}{t_0}\ C_2 \frac{\Gamma(a_2)}{\Gamma(b_2)}
\ _{3}{F}_{2}\left(\left. \begin{array}{ccc} 1 & 1 & a_2 \\ ~ & 2 & b_2\end{array}\right\vert {-\frac{Q^2}{t_0}}\right)\,;
\ee
the corresponding Adler function is
\be
\cA_2 (Q^2)=-Q^2 \frac{\partial \Pi_2(Q^2)}{\partial (Q^2)}=
\frac{Q^2}{t_0}\ C_2 \frac{\Gamma(a_2)}{\Gamma(b_2)}
\ _{3}{F}_{2}\left(\left. \begin{array}{ccc} 2 & 1 & a_2 \\ ~ & 2 & b_2\end{array}\right\vert {-\frac{Q^2}{t_0}}\right)\,,
\ee
and the {\it equivalent} $N=2$ spectral function is~\footnote{Notice the contrast with the predicted {\it equivalent} spectral function of the  Pad\'e approximant constructed with $\cM(0)$ and $\cM(-1)$ which is just a delta function.}:
\be
\frac{1}{\pi} \Imm\Pi_{2}(t) = \frac{t}{t_0}\; C_2 \;\Meijer{0,2}{0,2}{\frac{t}{t_0}}{0,1-a_2 \,; \,\relbar\!\relbar}{\relbar\!\relbar\, ;\, -1,1-b_2}\,.
\ee

\end{itemize}

We next propose to show the application of the Mellin-Barnes approximants discussed above to a non trivial example.

\section{\Large Mellin-Barnes-approximants (MBa) in QED at two loops.}
\setcounter{equation}{0}
\def\theequation{\arabic{section}.\arabic{equation}}

\noi 
We wish to test the techniques developed in the  previous section with a more complicated  example than the lowest order QED vacuum polarization discussed in ref.~\cite{EdeR17a}. We suggest to examine the case of the QED vacuum polarization at two loops.
The proper fourth order QED spectral function was first calculated by K\"{a}llen and Sabry in 1955~\cite{KS55} and later on in ref.~\cite{LdeR68}. It is given by the following expression:

With $m$ the lepton mass in the QED VP-loop and
\be
\delta= \sqrt{1-\frac{4m^2}{t}}\,,
\ee

{\setl
\bea\lbl{eq:KS}
\frac{1}{\pi}\Imm\Pi^{\rm QED}_{\rm 4th}(t) & = & \left(\frac{\alpha}{\pi} \right)^2\left\{
\delta\left(\frac{5}{8}-\frac{3}{8}\delta^2 -\left(\frac{1}{2}-\frac{1}{6} \delta^2\right)\log\left[64\frac{\delta^4}{(1-\delta^2)^3} \right]\right)\right.\nn \\
  & + & \left(\frac{11}{16}+\frac{11}{24}\delta^2 -\frac{7}{48}\delta^4
+\left(\frac{1}{2}+\frac{1}{3}\delta^2 -\frac{1}{6}\delta^4  \right)\log\left[\frac{(1+\delta)^3}{8\delta^2}\right]\right)\log\left[\frac{1+\delta}{1-\delta} \right] \nn  \\
& + & \left. 2\left(\frac{1}{2}+\frac{1}{3}\delta^2 -\frac{1}{6}\delta^4  \right)
\left(2\ \Li_2\left[\frac{1-\delta}{1+\delta} \right]+\Li_2 \left[-\frac{1-\delta}{1+\delta} \right] \right)\right\}\theta(t-4m^2)\,.
\eea}

\begin{figure}[!ht]
\begin{center}
\hspace*{-1cm}\includegraphics[width=0.50\textwidth]{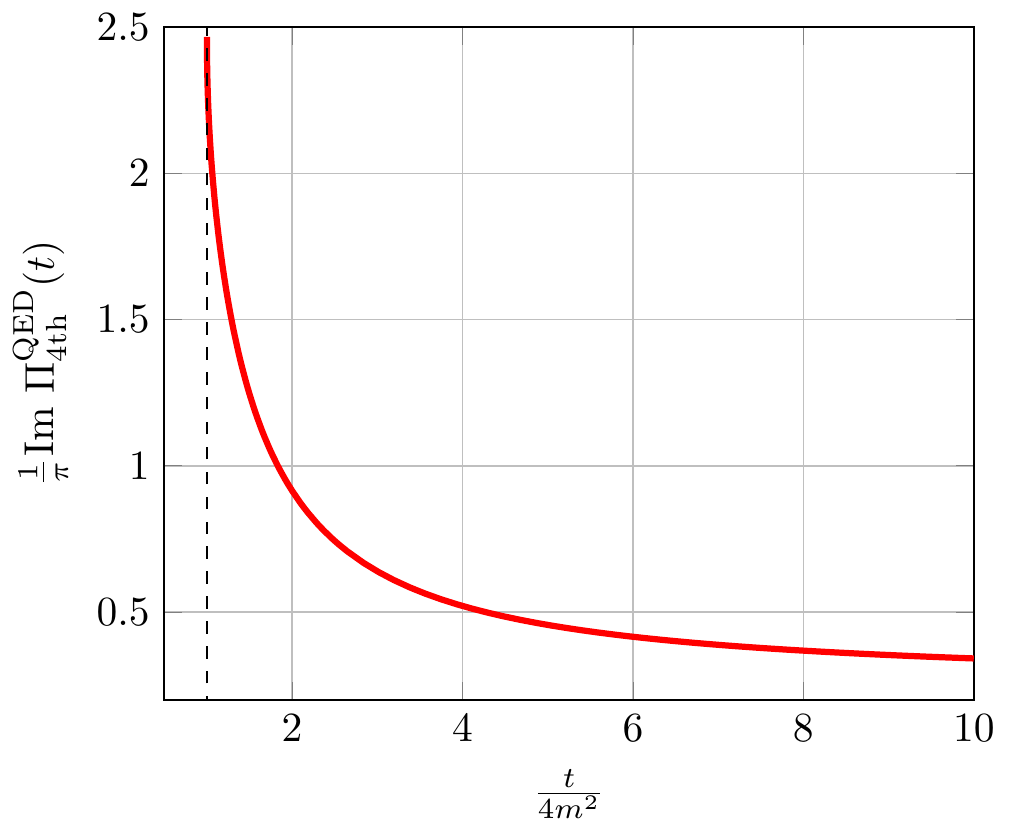} 
\bf\caption{\lbl{fig:spect4}}
\vspace*{0.25cm}
{\it Shape of the Spectral Function in  Eq.~\rf{eq:KS} in $\left(\frac{\alpha}{\pi} \right)^2$ units.}
\end{center}
\end{figure}

\noi
The asymptotic behaviours of this spectral function are

{\setl
\bea
\frac{1}{\pi}\Imm\Pi^{\rm QED}_{\rm 4th}(t) & \underset{{t\ra 4m^2}}{\thicksim} & \left(\frac{\alpha}{\pi} \right)^2\left\{\frac{\pi^2}{4}-2\sqrt{\frac{t}{4m^2}-1}+\frac{\pi^2}{6}\left(\frac{t}{4m^2}-1\right)+\cO\left[\left(\frac{t}{4m^2}-1\right)^{3/2}\right]\right\}\,, \lbl{eq:spect4thth}\\
\frac{1}{\pi}\Imm\Pi^{\rm QED}_{\rm 4th}(t) & \underset{{t\ra\infty}}{\thicksim} & \left(\frac{\alpha}{\pi} \right)^2\left\{\frac{1}{4}+\frac{3}{4}\frac{4m^2}{t}+\cO\left[\left(\frac{4m^2}{t} \right)^2 \log\left(\frac{t}{4m^2}\right)\right]\right\} \lbl{eq:spect4thinf}\,.
\eea}

\noi
Notice that the  behaviour at threshold $t\sim 4m^2$ is rather different  to the one at the one loop level~\cite{EdeR17a} and the shape of the spectral function, which is shown in Fig.~\rf{fig:spect4}, is also very different.
\begin{figure}[!ht]
\begin{center}
\hspace*{-1cm}\includegraphics[width=0.50\textwidth]{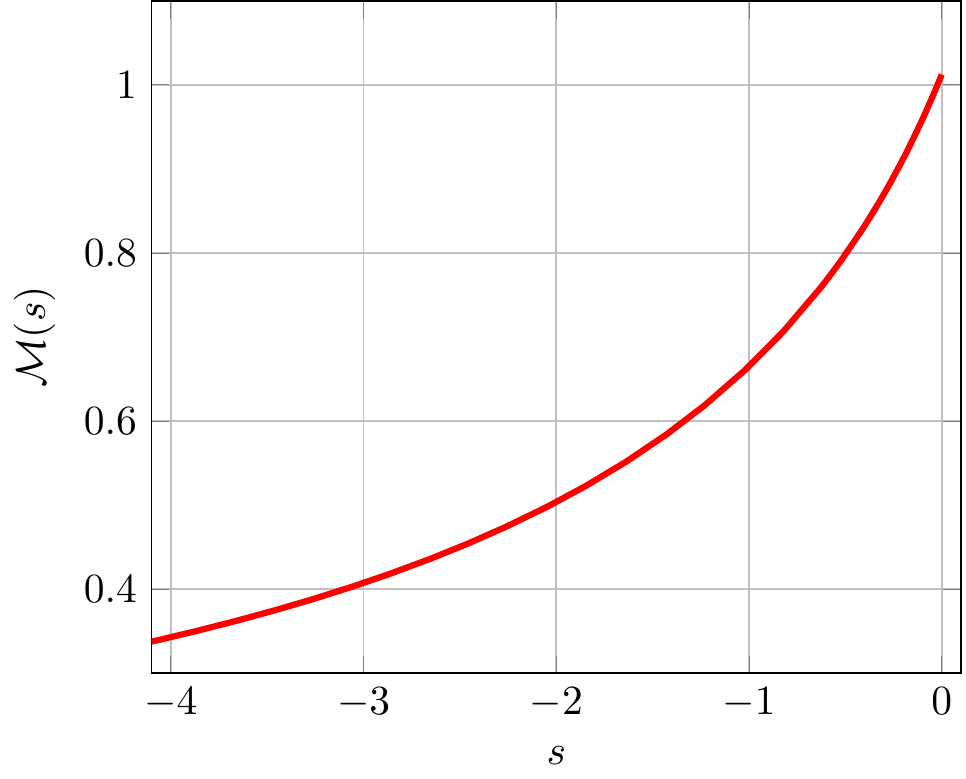} 
\bf\caption{\lbl{fig:melspec4}}
\vspace*{0.25cm}
{\it Shape of the Mellin Transform of the Spectral Function in  Eq.~\rf{eq:KS} in $\left(\frac{\alpha}{\pi} \right)^2$ units.}
\end{center}
\end{figure}

The shape of the Mellin transform of the 4th order spectral function in Eq.~\rf{eq:KS}  is shown in Fig.~\rf{fig:melspec4}. Like the Mellin transform  in QCD  it is also singular at $s=1$ but with a different residue
\be\lbl{eq:s1}
\cM^{\rm QED}_{\rm 4th}(s)\underset{{s\ra 1}}{\thicksim} \left(\frac{\alpha}{\pi}\right)^2 \frac{1}{4}\frac{1}{1-s}\,,
\ee
and shares with QCD the property of being a monotonously increasing function from   $s=-\infty$ to $s<1$.

The real part of the fourth order vacuum polarization in QED is also known analytically~\cite{KS55}.  It is a rather complicated expression and, therefore, it is a good test to see how well it is approximated by the successive GH-Functions in Eq.~\rf{eq:GHF}. The shape   of the  $\Pi^{\rm QED}_{\rm 4th}(Q^2)$ function  in the Euclidean is shown in Fig.~\rf{fig:pi4E}.

We shall discuss this 4th order QED example in a way as close as possible to the QCD case which we shall later be confronted with. Therefore,  the input will be the successive values of the moments of the spectral function, i.e. of the derivatives of $\Pi_{\rm 4th}^{\rm QED}(Q^2)$ at $Q^2 =0$.

\begin{figure}[!ht]
\begin{center}
\hspace*{-1cm}\includegraphics[width=0.50\textwidth]{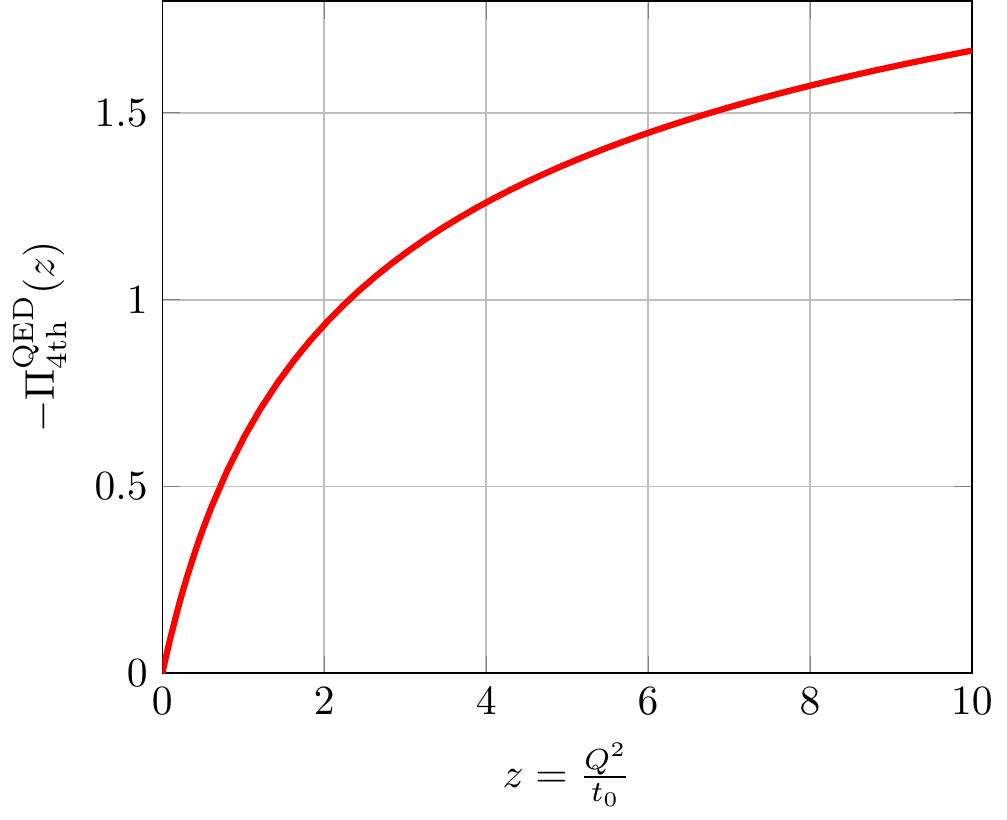} 
\bf\caption{\lbl{fig:pi4E}}
\vspace*{0.25cm}
{\it Shape of the 4th order QED vacuum polarization function in the Euclidean \\  $\left(\frac{\alpha}{\pi} \right)^2$ units.}
\end{center}
\end{figure}

\noi
The first few Mellin moments

\be
\cM^{\rm QED}_{\rm 4th}(s)\equiv\int_{4m^2}^\infty \frac{dt}{t}\left(\frac{t}{4m^2}\right)^{s-1}\frac{1}{\pi}\Imm\Pi^{\rm QED}_{\rm 4th}(t)\,,
\ee
for $s=0,-1,-2,-3,-4,-5$,  in units of $\left(\frac{\alpha}{\pi} \right)^2$ are tabulated below in Table~\rf{table:t1}. 

\begin{table*}[h]
\caption[Results]{ $\cM(s)$ Moments in units of $\left(\frac{\alpha}{\pi} \right)^2$. } 
\lbl{table:t1}
\begin{center}
\begin{tabular}{|c|c|c|} \hline \hline {\bf Moment} & {\bf Exact result} & {\bf Numerical value}
\\ 
\hline \hline
$\cM(0)$ &  $82/81$ & $1.012356796$ 
\\
$\cM(-1)$ & $449/675$ & $0.665185185$ \\
$\cM(-2)$ & $249916/496125$ & $0.503735936$ \\
$\cM(-3)$ & $51986/127575$ & $0.407493631$  \\
$\cM(-4)$ & $432385216/1260653625$ & $ 0.342984946$  \\
$\cM(-5)$ & $5415247216/18261468225$ & $0.296539531$  \\
\hline\hline
\end{tabular}
\end{center}
\end{table*} 

\subsection{\large Successive Approximations to $\cM^{\rm QED}_{\rm 4th}(s)$, $\Pi^{\rm QED}_{\rm 4th}(Q^2)$ and $a_{\mu}^{\rm VP}$.}

\noi
We can now proceed to the construction of a successive set of MBa's to $\cM^{\rm QED}_{\rm 4th}(s)$ of the type shown  in Eq.~\rf{eq:marichevend}  and to the evaluation of the corresponding GH-function  approximation to $\Pi^{\rm QED}_{\rm 4th}(Q^2)$ of the type shown in Eq.~\rf{eq:GHF}. At each approximation step we shall then evaluate  the corresponding  contribution to the anomalous magnetic moment of a fermion of mass $m$ induced by the 4th order vacuum polarization generated by the same fermion (see the corresponding  Feynman diagrams in Fig.~\rf{fig:QED4}),
and compare it with the exact result which is known analytically~\cite{MR69}:

{\setl
\bea\lbl{eq:MiRe}
a^{\rm VP}_{\mu} & = & \left(\frac{\alpha}{\pi} \right)^3 \left\{\frac{673}{108}-\frac{41}{81}\pi^2
-\frac{4}{9}\pi^2 \log(2)-\frac{4}{9}\pi^2 \log^{2}(2) +\frac{4}{9}\log^{4}(2) -\frac{7}{270}\pi^4 \right.\nn \\
 &  & \left. +\frac{13}{18}\zeta(3)+\frac{32}{3}{\rm PolyLog}\left[4\,, \frac{1}{2} \right]\right\} =\left(\frac{\alpha}{\pi} \right)^3  0.0528707\,.
\eea}

\begin{figure}[!ht]
\begin{center}
\includegraphics[width=0.70\textwidth]
{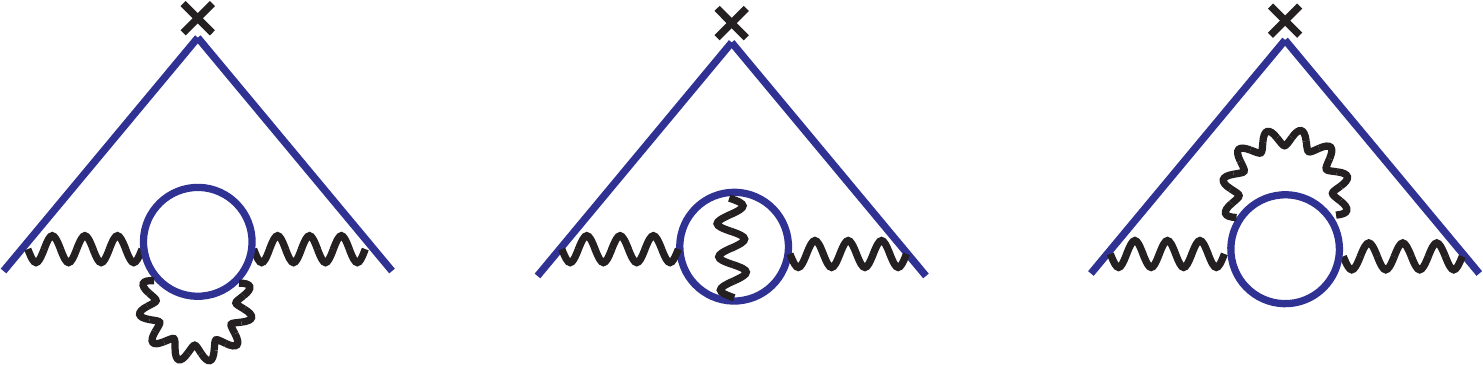} 
\bf\caption{\lbl{fig:QED4}}
{\it Feynman diagrams contributing to the muon anomaly in Eq.~\rf{eq:MiRe}.}
\vspace*{0.25cm}
 
\end{center}
\end{figure}

\noi 
The result in Eq.~\rf{eq:MiRe} is a rather complicated expression involving higher transcendental numbers with important numerical cancellations among the different terms and, therefore, it should provide a good test. We want to investigate how well  we reproduce this exact result using the Mellin-Barnes integral representation in Eq.~\rf{eq:MBamuF} which, when adapted to this case, reads 	as follows:
\be\lbl{eq:MBamuNQED}
a^{\rm VP}(N) =   \left(\frac{\alpha}{\pi}\right) \frac{1}{2}\frac{1}{2\pi}\int\limits_{-\infty}^{+\infty}d\tau\ e^{-i\tau\log 4}\  \cF\left(\frac{1}{2}-i\tau \right)\ \cM_{N}\left(\frac{1}{2}-i\tau \right)\,,
\ee
with  $\cM_{N}(s)$ the successive   Mellin approximants.

\subsubsection{\normalsize The $N=1$ MBa.}

\noi
This corresponds to the case where we only know $\cM_{\rm 4th}^{\rm QED}(0)$. Following Eq.~\rf{eq:marichevend} we are instructed to consider as a first Mellin approximant:

\be\lbl{eq:1MQED}
\cM^{\rm QED}_{\rm 4th}(s)\Ra \cM_{1}(s)=  C_{1}\frac{\Gamma(a-s)}{\Gamma(b-s)}\,,
\ee
which must be singular at $s=1$. This fixes the $a$ parameter to $a=1$ and the overall normalization to
\be
C_{1}= \left(\frac{\alpha}{\pi}\right)^2 \frac{1}{4}\Gamma(b-1)\,,
\ee
so as to reproduce the leading singularity when $s\ra 1$.
Matching $\cM_{1}(s)$ at $s=0$ with the numerical value of  $\cM_{\rm 4th}^{\rm QED}(0)$ in Table~\rf{table:t1} fixes the $b$ parameter to
\be
b=1.24695122\,.
\ee

We can then perform the corresponding integral in 
Eq.~\rf{eq:MBamuNQED} which gives as a result for the first $N=1$ approximant:
\be\lbl{eq:QEDN1}
a^{\rm VP} (N=1)=\left(\frac{\alpha}{\pi} \right)^3 \times 0.0500007\,.
\ee
It reproduces the Mignaco-Remiddi exact  result in Eq.~\rf{eq:MiRe} to an accuracy of 5\%.

\subsubsection{\normalsize The $N=2$ MBa.}

\noi
This corresponds to the case where  we know the slope and curvature of $\Pi_{\rm 4th}^{\rm QED}(Q^2)$ at $Q^2 =0$, i.e. $\cM_{\rm 4th}^{\rm QED}(0)$ and $\cM_{\rm 4th}^{\rm QED}(-1)$. This information is similar to  that already available from LQCD~\footnote{See refs.~\cite{Burger14,Ch16,Lellouch16} and references therein.}. We shall therefore  discuss it in detail.  

The Mellin approximant in this case has two parameters $a$ and $b$:
\be\lbl{eq:2MQED}
\cM^{\rm QED}_{\rm 4th}(s)\Ra \cM_{2}(s)=  C_{2}\frac{\Gamma(1-s)}{\Gamma(2-s)}\frac{\Gamma(a-s)}{\Gamma(b-s)}\,,
\ee
and the leading short-distance constraint fixes the overall normalization to
\be
C_{2}=\left(\frac{\alpha}{\pi}\right)^2 \frac{1}{4} \frac{\Gamma(b-1)}{\Gamma(a-1)}\,,
\ee
with the parameters $a$ and $b$ fixed by the two matching equations:
\be
\frac{1}{4}\frac{a-1}{b-1}=\cM_{\rm 4th}^{\rm QED}(0)\quad\annd\quad
\frac{1}{8}\frac{a}{b}\frac{a-1}{b-1}=\cM_{\rm 4th}^{\rm QED}(-1)\,,
\ee
or equivalently

{\setl
\bea
\frac{1}{4}\frac{a-1}{b-1} & = & \cM_{\rm 4th}^{\rm QED}(0)\\
\frac{1}{2}\frac{a}{b} & = & \frac{\cM_{\rm 4th}^{\rm QED}(-1)}{\cM_{\rm 4th}^{\rm QED}(0)}\,.
\eea}

\begin{figure}[!ht]
\begin{center}
\hspace*{-1cm}\includegraphics[width=0.50\textwidth]{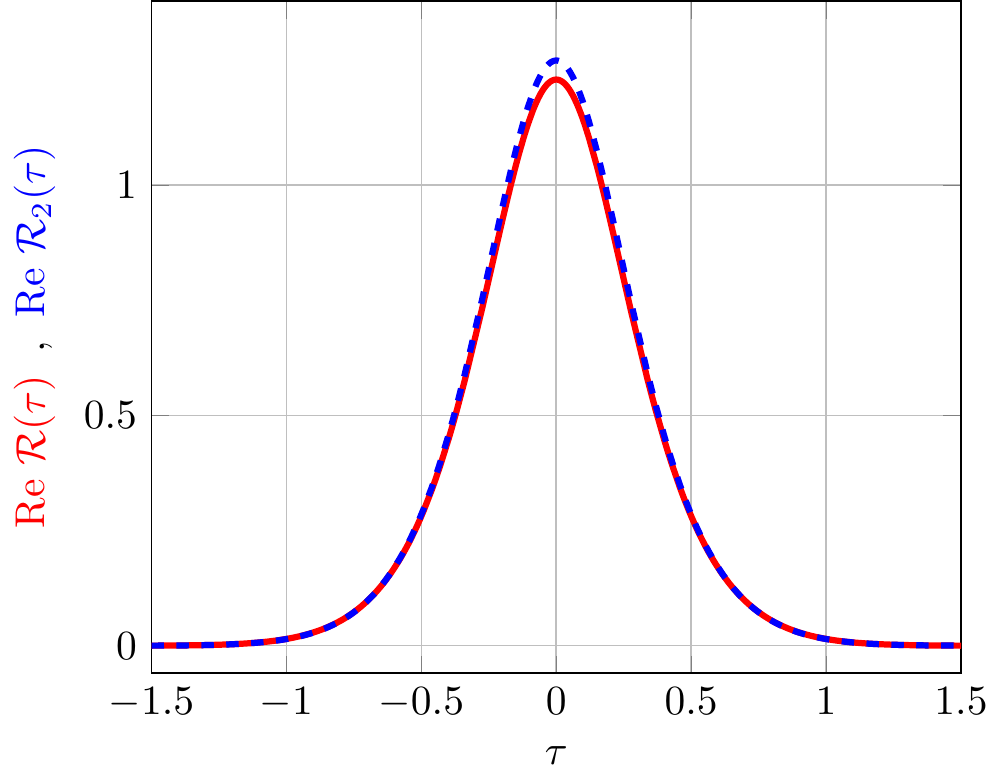} 
\bf\caption{\lbl{fig:intgrand}}
\vspace*{0.25cm}
{\it Plot of the real part of the integrand $\cR_{2}(\tau)$ in Eq.~\rf{eq:fourier2}:\\ the red curve corresponds to inserting the exact $\cM^{\rm QED}_{\rm 4th}\left(\frac{1}{2}-i\tau \right)$ in the integrand,\\ the dashed blue curve to  inserting  the approximation  $\cM_{2}\left(\frac{1}{2}-i\tau \right)$.} 
\end{center}
\end{figure}

\begin{figure}[!ht]
\begin{center}
\hspace*{-1cm}\includegraphics[width=0.50\textwidth]{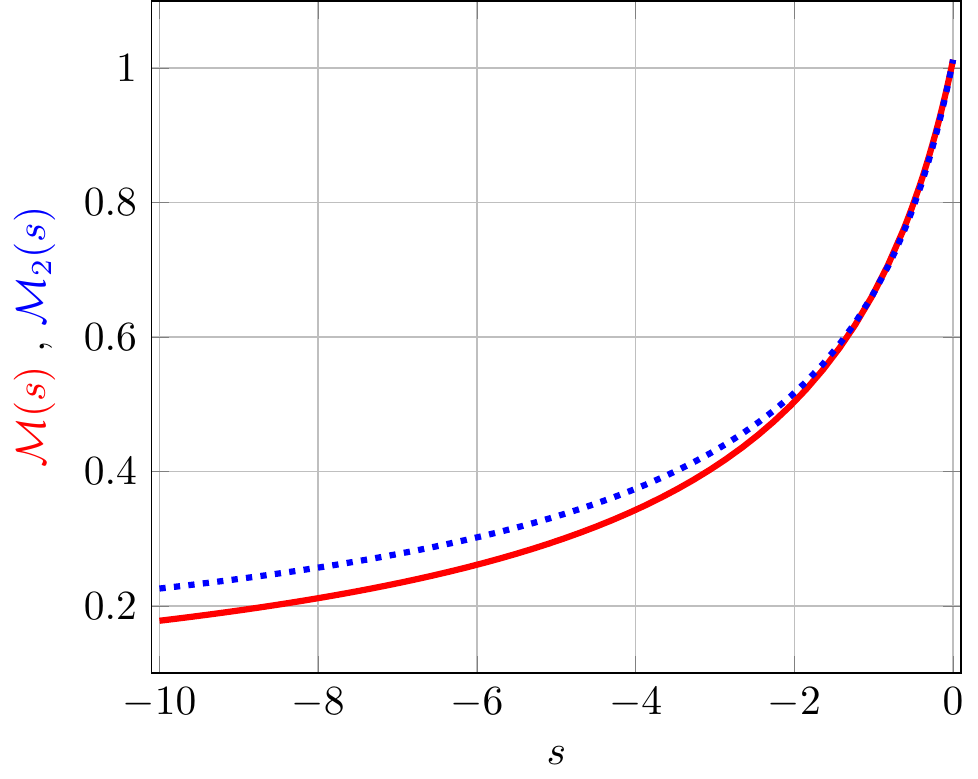} 
\bf\caption{\lbl{fig:melexM2}}
\vspace*{0.25cm}
{\it The red curve is the Mellin Transform of the Spectral Function in  Eq.~\rf{eq:KS}.\\ 
The dotted blue curve is the $N=2$ Mellin approximant in Eq.~\rf{eq:2MQED}.\\  Both curves are shown in $\left(\frac{\alpha}{\pi} \right)^2$ units.}
\end{center}
\end{figure}

\noi
Inserting the numerical values in Table~\rf{table:t1} for $\cM_{\rm 4th}^{\rm QED}(0)$ and $\cM_{\rm 4th}^{\rm QED}(-1)$ results in the values
\be\lbl{eq:pars2}
a=1.46508\quad\annd\quad b=1.11485\,.
\ee
With these parameter values inserted in $\cM_2 (s)$ in Eq.~\rf{eq:2MQED}, and performing the corresponding integral
\be\lbl{eq:fourier2}
a_{\mu}^{\rm VP}(N=2)  = \left(\frac{\alpha}{\pi}\right)\frac{1}{2} \frac{1}{2\pi}\int\limits_{-\infty}^{+\infty}d\tau\  \underbrace{e^{-i\tau \log 4}\ \cF\left(\frac{1}{2}-i\tau\right)\ \cM_{2}\left(\frac{1}{2}-i \tau \right)}_{\cR_{2}(\tau)}\,,
\ee
 gives the result
\be
a^{\rm VP} (N=2)=\left(\frac{\alpha}{\pi} \right)^3 \times 0.0531447\,,
\ee
which reproduces the Mignaco-Remiddi  result in Eq.~\rf{eq:MiRe} to an accuracy of 0.5\%, a significant improvement with respect to the $N=1$ approximant. Figure~\rf{fig:intgrand} shows the behaviour of the real part of the  integrand $\cR_{2}(\tau)$ in Eq.~\rf{eq:fourier2} as a function of $\tau$, where the red curve is the one when one inserts the exact Mellin transform $\cM^{\rm QED}_{\rm 4th}\left(\frac{1}{2}-i\tau \right)$ in the integrand and the dashed blue curve the one associated to the $N=2$ approximation. Already at this level of approximation the agreement between both integrands is quite impressive.

\begin{figure}[!ht]
\begin{center}
\hspace*{-1cm}\includegraphics[width=0.50\textwidth]{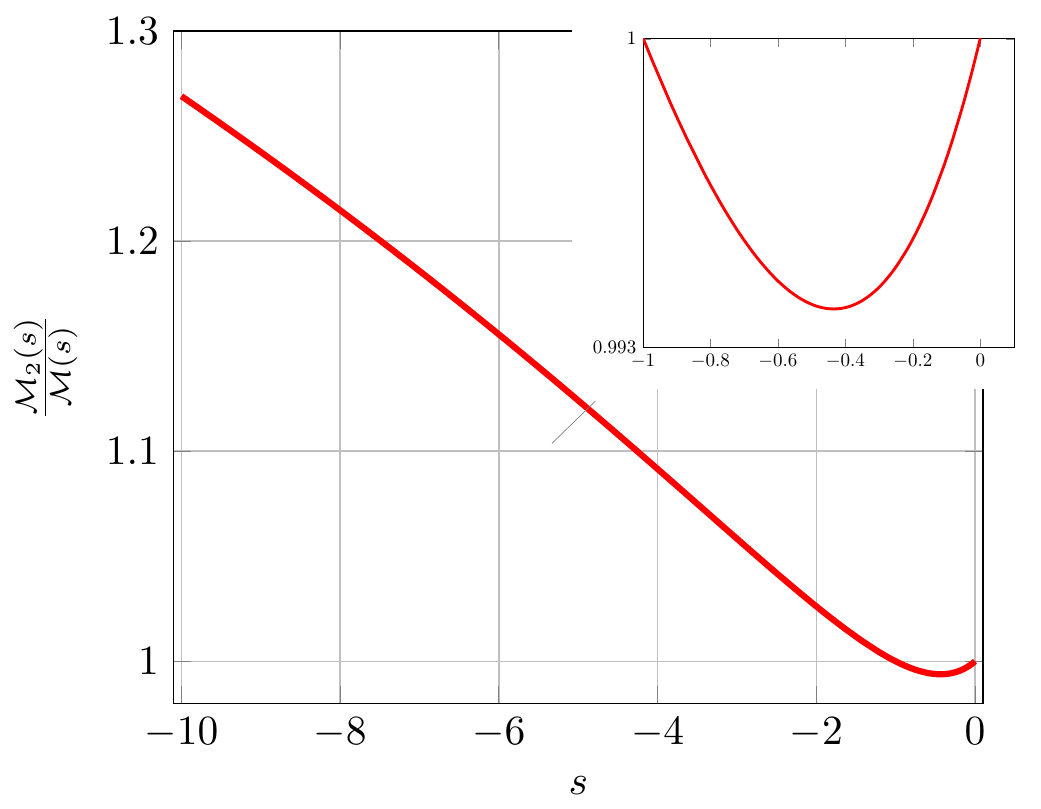} 
\bf\caption{\lbl{fig:R2}}
\vspace*{0.25cm}
{\it Plots of the ratio $\frac{\cM_{2}(s)}{\cM(s)}$ versus $s$. Notice the scale of the plots. }
\end{center}
\end{figure}

At this stage it is also interesting to compare the exact Mellin transform shown in Fig~\rf{fig:melspec4} with the one corresponding to the $N=2$ approximation. This is shown in Fig.~\rf{fig:melexM2} where the blue dotted curve is the $N=2$ approximation. The agreement of the two curves down to $s\simeq -3$ is quite remarkable. In order to see the difference between these two curves we show in Fig.~\rf{fig:R2} the plot of their ratio. The $\cM_{2}(s)/\cM(s)$ ratio turns out to be greater than one everywhere, except in the interval $-1\le s \le 0$. This is why  the $N=2$ result approaches the exact value of the anomaly from above. The quality of the interpolation between $s=0$ and $s=-1$ provided by the $N=2$ approximation is shown at the right in Fig.~\rf{fig:R2}. Notice the scale in the figure, e.g. the value at the minimum of the ratio shown in this figure is $0.9937$ compared to one.

\begin{figure}[!ht]
\begin{center}
\hspace*{-1cm}\includegraphics[width=0.50\textwidth]{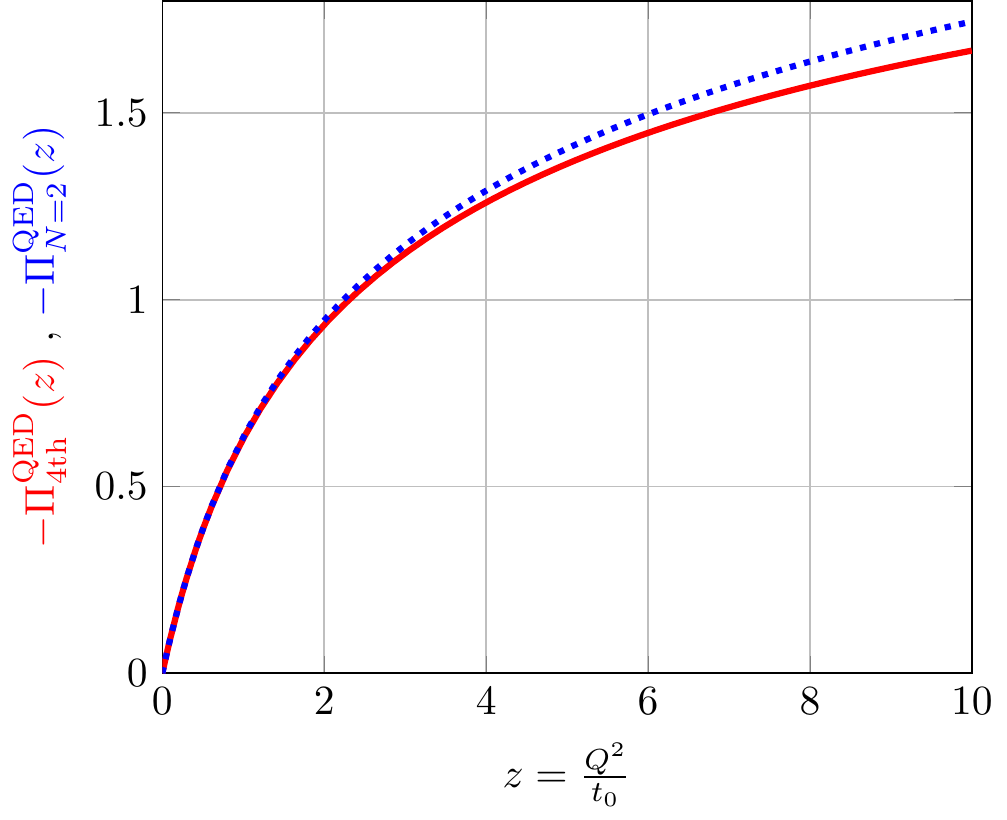} 
\bf\caption{\lbl{fig:RPi22}}
\vspace*{0.25cm}
{\it The red curve is the exact 4th order QED VP-function.\\ 
The dotted blue curve is the $N=2$ approximant.\\  Both curves are shown in $\left(\frac{\alpha}{\pi} \right)^2$ units.}
\end{center}
\end{figure}

According to Eq.~\rf{eq:GHF}, the $N=2$ GH-function approximant to $\Pi_{\rm 4th}^{\rm QED}(Q^2)$ is given by the expression ($z\equiv \frac{Q^2}{4m^2}$):
\be\lbl{eq:GHFN2}
\Pi_{\rm 4th}^{\rm QED}(Q^2)\Ra \Pi_{(N=2)}^{\rm QED}(Q^2) = 
\left(\frac{\alpha}{\pi} \right)^2 \ (-z)\frac{1}{4}\frac{a-1}{b-1}\ _{3}{F}_{2}\left(\left. \begin{array}{ccc} 1 & 1 & a \\ ~ & 2 & b\end{array}\right\vert {-z}\right)\,,
\ee

\noi
where $\ _{3}{F}_{2}\left(\left. \begin{array}{ccc} 1 & 1 & a \\ ~ & 2 & b\end{array}\right\vert {-}\right)$ is the GH-Function defined by the series:
\be
 _{3}{F}_{2}\left(\left. \begin{array}{ccc} 1 & 1 & a \\ ~ & 2 & b\end{array}\right\vert {-z}\right)=\sum_{n=0}^{\infty}\frac{(1)_n (1)_n (a)_n }{(2)_n (b)_n}\frac{(-z)^n}{n!}\,,
\ee
and $a$ and $b$ have the values given in Eq.~\rf{eq:MBamuF}. 
Figure \rf{fig:RPi22} shows how well the MBa $\Pi_{(N=2)}^{\rm QED}(Q^2)$ (blue curve) does when compared to the exact function (red curve). From this comparison, one can qualitatively understand why the $N=2$ approximation already reproduces the exact value of $a^{\rm VP}$ in Eq.~\rf{eq:MiRe} at the $0.5\%$ level.

The {\it equivalent} spectral function corresponding to the $N=2$ approximation is given by the Meijer's G-Function:
\be
\frac{1}{\pi} \Imm\Pi_{2}(t) = \frac{t}{t_0}\; \left(\frac{\alpha}{\pi} \right)^2 \frac{1}{4}  \;\Meijer{0,2}{0,2}{\frac{t}{t_0}}{0,1-a \,; \,\relbar\!\relbar}{\relbar\!\relbar\, ;\, -1,1-b}\,,
\ee
and its shape,  compared to the exact spectral function, is shown in Fig.~\rf{fig:Pi22}. Notice that the {\it equivalent} spectral function corresponding to the unique Pad\'e approximant constructed with $\cM_{\rm 4th}^{\rm QED}(0)$ and $\cM_{\rm 4th}^{\rm QED}(-1)$ would be just a delta function.
\begin{figure}[!ht]
\begin{center}
\hspace*{-1cm}\includegraphics[width=0.50\textwidth]{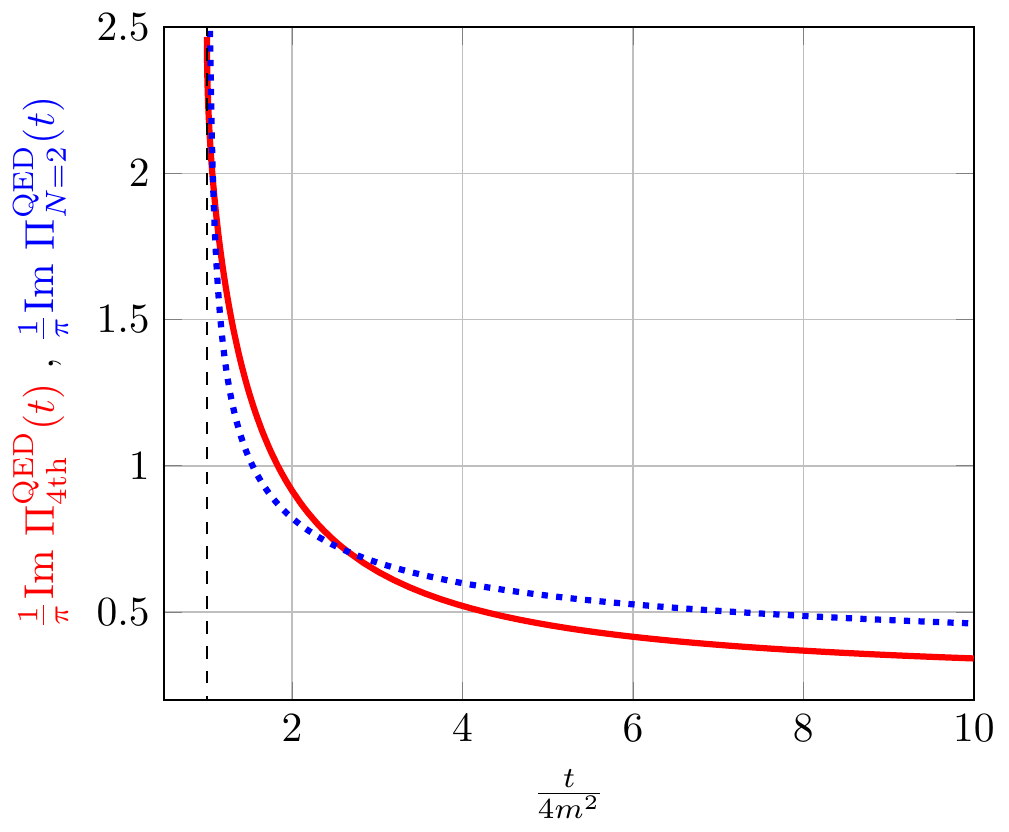} 
\bf\caption{\lbl{fig:Pi22}}
\vspace*{0.25cm}
{\it The red curve is the exact 4th order QED spectral function.\\ 
The dotted blue curve is the $N=2$ approximant.\\  Both curves are shown in $\left(\frac{\alpha}{\pi} \right)^2$ units.}
\end{center}
\end{figure}

\subsubsection{\normalsize The $N=3$ MBa.}

\noi
This corresponds to the Mellin approximant
\be\lbl{eq:3MQED}
\cM^{\rm QED}_{\rm 4th}(s)\Ra \cM_{3}(s)=  C_{3}\frac{\Gamma(1-s)\Gamma(a_1 -s)}{\Gamma(b_1 -s)\Gamma(b_2 -s)}\,,
\ee
with
\be
C_{3}=\left(\frac{\alpha}{\pi}\right)^2 \frac{1}{4} \frac{\Gamma(b_1 -1)\Gamma(b_2 -1)}{\Gamma(a_1 -1)}\,,
\ee
and the three parameters $a_1$, $a_2$ and $b_1$  fixed by  matching $\cM_{3}(s)$ to the values of the three moments $\cM^{\rm QED}_{\rm 4th}(0)$, $\cM^{\rm QED}_{\rm 4th}(-1)$, and $\cM^{\rm QED}_{\rm 4th}(-2)$. The matching equations in this case are:

{\setl
\bea
\left(\frac{\alpha}{\pi}\right)^2 \frac{1}{4}\frac{1}{b_{1}-1}(a_{1}-1)\frac{1}{b_{2}-1} & = & \cM^{\rm QED}_{\rm 4th}(0)\,, \\
\frac{1}{b_{1}}a_{1}\frac{1}{b_{2}} & = & \frac{\cM^{\rm QED}_{\rm 4th}(-1)}{\cM^{\rm QED}_{\rm 4th}(0)}\,,\\
2\frac{1}{b_{1}+1}(a_{1}+1)\frac{1}{b_{2}+1} & = & \frac{\cM^{\rm QED}_{\rm 4th}(-2)}{\cM^{\rm QED}_{\rm 4th}(-1)}\,,
\eea}

\noi
which results in the values:
\be
a_{1} = 
2.528554853\,,\quad
b_{1} =  1.163614902\,,\quad
b_{2}  = 3.307115556\,,
\ee
or the equivalent solution with $b_{1}\rightleftharpoons b_{2}$. With these values inserted in $\cM_3 (s)$ in Eq.~\rf{eq:2MQED}, and performing the corresponding integral in 
Eq.~\rf{eq:MBamuNQED} gives the result
\be
a^{\rm VP} (N=3)=\left(\frac{\alpha}{\pi} \right)^3 \times 0.0528678\,,
\ee
which now reproduces the Mignaco-Remiddi  result in Eq.~\rf{eq:MiRe} to the remarkable accuracy of 0.004\%.

\begin{figure}[!ht]
\begin{center}
\hspace*{-1cm}\includegraphics[width=0.50\textwidth]{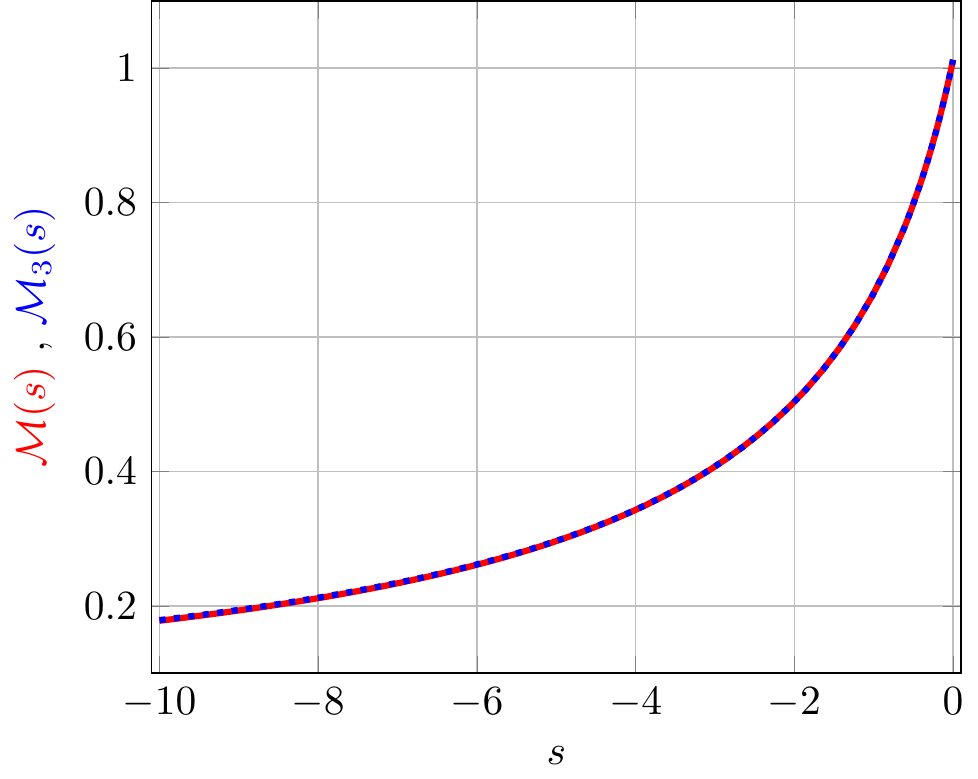} 
\bf\caption{\lbl{fig:Mel11012}}
\vspace*{0.25cm}
{\it The red curve is the Mellin Transform of the exact Spectral Function.\\ 
The dashed blue curve is the $N=3$ Mellin approximant.  Both curves are shown in $\left(\frac{\alpha}{\pi} \right)^2$ units.}
\end{center}
\end{figure}
\begin{figure}[!ht]
\begin{center}
\hspace*{-1cm} \includegraphics[width=0.50\textwidth]{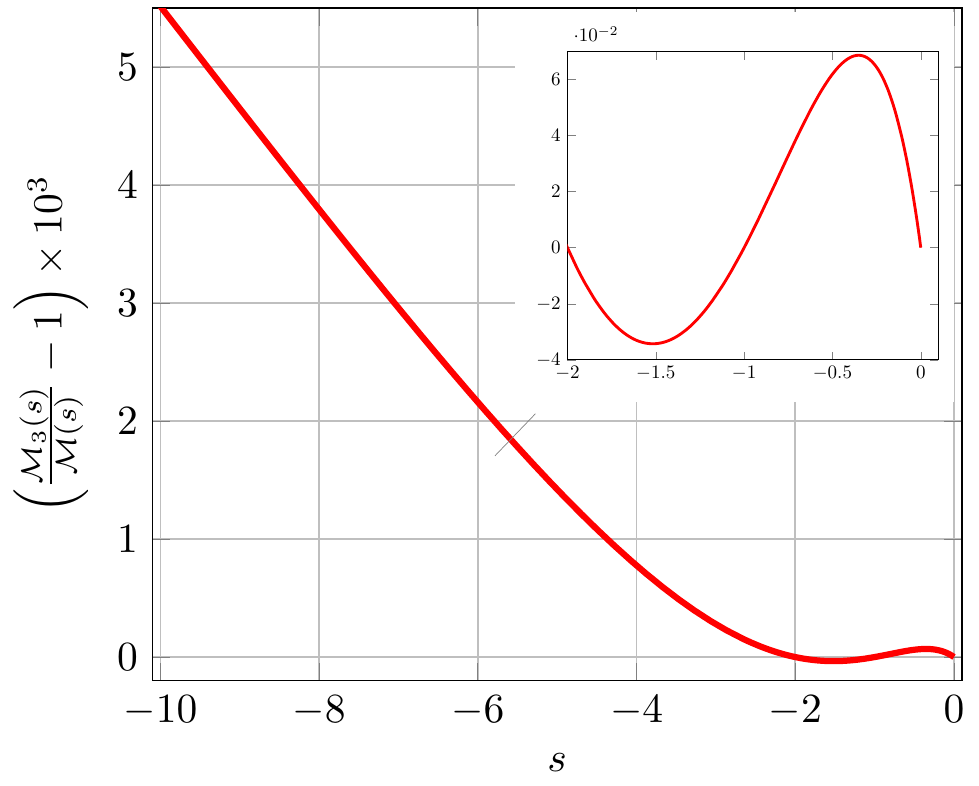} 
\bf\caption{\lbl{fig:R33S}}
\vspace*{0.25cm}
{\it Plots of the ratio $\frac{\cM_{3}(s)}{\cM(s)}$ versus $s$. Notice the vertical scales of these plots. }
\end{center}
\end{figure}

As an illustration of the quality of the approximation,  we show in Fig.~\rf{fig:Mel11012} the Mellin transform of the $N=3$ approximation (the blue dashed curve) compared to the exact Mellin transform (the red curve). At the scale of the figure it is practicably impossible to see the difference.  In order to see that, we show plots of the ratio $\cM_{3}(s)/\cM(s)$ in Fig.~\rf{fig:R33S}. Notice the scale in the left plot of Fig.~\rf{fig:R33S} as compared to the one in Fig.~\rf{fig:R2} and the improvement in the figure at the right which is plotted at the same scale as Fig.~\rf{fig:R2}.

An accuracy of 0.004\% is already much beyond what is required of the HVP contribution to the muon anomaly in QCD, but for the sake of testing the approximation procedure that we are advocating, let us try further possible improvements.

\subsubsection{\normalsize The $N=4$ MBa.}

\noi
The $N=4$ approximant is 
\be\lbl{eq:4MQED}
\cM^{\rm QED}_{\rm 4th}(s)\Ra \cM_{4}(s)=  C_{4}
\frac{\Gamma(1-s)\Gamma(a_1 -s)\Gamma(a_2 -s)}{\Gamma(2-s)\Gamma(b_1 -s)\Gamma(b_2 -s)}\,,
\ee
with
\be
C_{4}=\left(\frac{\alpha}{\pi}\right)^2 \frac{1}{4} \frac{\Gamma(b_1 -1)\Gamma(b_2 -1)}{\Gamma(a_1 -1)\Gamma(a_2 -1)}\,,
\ee
and the four parameters $a_1$, $a_2$, $b_1$ and $b_2$ solutions of the matching equations:

{\setl
\bea
\frac{1}{4}\frac{a_1-1}{b_1-1}\frac{a_2-1}{b_2-1} & = & \cM^{\rm QED}_{\rm 4th}(0)\,, \\
\frac{1}{2}\frac{a_1}{b_1}\frac{a_2}{b_2} & = & \frac{\cM^{\rm QED}_{\rm 4th}(-1)}{\cM^{\rm QED}_{\rm 4th}(0)}\,,\\
\frac{2}{3} \frac{(a_1 +1)}{(b_1 +1)} \frac{(a_2 +1)}{(b_2 +1)} & = & \frac{\cM^{\rm QED}_{\rm 4th}(-2)}{\cM^{\rm QED}_{\rm 4th}(-1)}\,, \\
\frac{3}{4} \frac{(a_1 +2)}{(b_1 +2)} \frac{(a_2 +2)}{(b_2 +2)} & = & \frac{\cM^{\rm QED}_{\rm 4th}(-3)}{\cM^{\rm QED}_{\rm 4th}(-2)}\,,
\eea}

\noi
which give, as an acceptable solution, the values:
\be
 a_ 1= 
2.829673582\,,\quad b_1 = 3.528046148\,,\quad a_2 = 1.902891314\,,\quad  b_2 = 1.161374634\,, 
\ee
or the equivalent solution with $a_{1}\rightleftharpoons a_{2}$ and $b_{1}\rightleftharpoons b_{2}$.

The corresponding prediction for the muon anomaly is
\be
a_{\mu}^{\rm VP}(N=4)= \left(\frac{\alpha}{\pi} \right)^3 0.0528711\,,
\ee
which reproduces the exact value at the level of 0.00075\%, practically the exact result.

It seems fair to conclude from these examples that the successive use of MBa of the Marichev class in Eq.~\rf{eq:marichevend} is an excellent method to approach, rather quickly in this case, the exact result with an excellent accuracy. The question which, however, arises is: {\it how far can one go?}. The exact Mellin transform of the QED fourth order spectral function, contrary to the second order one discussed in ref.~\cite{EdeR17a}, is expected to be a much more complicated expression than just a simple {\it standard product} of the Marichev class in Eq.~\rf{eq:marichevend}. Therefore, {\it a priori}, one expects these approximations to break at some $N$-level where no acceptable solutions exist any longer.  Let us then proceed  to examine what happens when one tries higher $N$-approximants of a single  {\it standard product}.

\subsubsection{\normalsize The $N=5$ MBa.}

\noi
The $N=5$ Mellin approximant is 
\be\lbl{eq:5MQED}
\cM^{\rm QED}_{\rm 4th}(s)\Ra \cM_{5}(s)=  C_{5}
\frac{\Gamma(1-s)\Gamma(a_1 -s)\Gamma(a_2 -s)}{\Gamma(b_1 -s)\Gamma(b_2 -s)\Gamma(b_3 -s)}\,,
\ee
with
\be
C_{5}=\left(\frac{\alpha}{\pi}\right)^2 \frac{1}{4} \frac{\Gamma(b_1 -1)\Gamma(b_2 -1)\Gamma(b_3 -1)}{\Gamma(a_1 -1)\Gamma(a_2 -1)}\,,
\ee
and the parameters $a_1$, $a_2$, $b_1$, $b_2$, $b_3$ solutions of the matching equations:

{\setl
\bea
\frac{1}{4}\frac{a_1 -1}{b_1 -1}\frac{a_2 -1}{b_2 -1}\frac{1}{b_3-1} & = & \cM^{\rm QED}_{\rm 4th}(0)\,, \\
\frac{a_1}{b_1}\frac{a_2}{b_2}\frac{1}{b_3} & = & \frac{\cM^{\rm QED}_{\rm 4th}(-1)}{\cM^{\rm QED}_{\rm 4th}(0)}\,,\\
2 \frac{a_1 +1}{b_1 +1} \frac{a_2 +1}{b_2 +1}\frac{1}{b_3 +1} & = & \frac{\cM^{\rm QED}_{\rm 4th}(-2)}{\cM^{\rm QED}_{\rm 4th}(-1)}\,, \\
3 \frac{a_1 +2}{b_1 +2} \frac{a_2 +2}{b_2 +2}\frac{1}{b_3 +2} & = & \frac{\cM^{\rm QED}_{\rm 4th}(-3)}{\cM^{\rm QED}_{\rm 4th}(-2)}\,, \\
4 \frac{a_1 +3}{b_1 +3} \frac{a_2 +3}{b_2 +3}\frac{1}{b_3 +3} & = & \frac{\cM^{\rm QED}_{\rm 4th}(-4)}{\cM^{\rm QED}_{\rm 4th}(-3)}\,.
\eea}

\noi
There are still acceptable solutions to this system of polynomial equations with the values:
\be
b_1 = 1.16249580\,, a_1 = 4.111523616\,, b_2 = 4.354959443\,, a_2 = 2.360299888\,, b_3 = 
2.917297589\,,  
\ee
and the permutations of $a_1$, $a_2$ and $b_1$, $b_2$, $b_3$ which give equivalent solutions.
The corresponding prediction for the muon anomaly is now
\be
a_{\mu}^{\rm VP}(N=5)= \left(\frac{\alpha}{\pi} \right)^3 0.0528706\,,
\ee
which reproduces the exact value at the level of 0.00018\%, still an improvement with respect to the $N=4$ Approximation!
 
 This is, however,  the  best one can do in the two loop QED case with single Mellin approximants of the type shown in Eq.~\rf{eq:marichevend}. Indeed, if one tries to improve with a $N=6$ approximant of this type, one finds that all the solutions for the parameters $a_1$, $a_2$, $a_3$, $b_1$, $b_2$, $b_3$  from the matching equations bring in complex  numbers with real parts which are inside of the {\it fundamental strip}, in contradiction with the initial requirements for an acceptable solution that we imposed. This is the signal that, in our example, single Marichev-like approximants  break down at a critical $N$-level where the function $\Pi_{\rm 4th}^{\rm QED}(Q^2)$ cannot be approximated any longer with just one GH-Function. It is possible, however, to extend the class of   approximants to  {\it superpositions of standard products} as indicated in Eq.~\rf{eq:marichevend} and in fact this is what we shall do in the case of QCD.

From the previous analysis we conclude that, in the case of the QED fourth order vacuum polarization, the best prediction we can make with single Marichev-like MBa's is an average of the $N=4$ and $N=5$ approximants with an error estimated from the deviation of this average  to the $N=4$ and $N=5$ results i.e.,
\be
 a_{\mu}^{\rm VP}(\rm QED~4th~order)=\left(\frac{\alpha}{\pi} \right)^3 (0.0528709\pm 0.0000003)\,.
\ee
This  is already an excellent prediction when  compared to the exact result in Eq.~\rf{eq:MiRe}.

\section{\Large Test of  MBa  with experimental HVP Moments.}
\setcounter{equation}{0}
\def\theequation{\arabic{section}.\arabic{equation}}

\noi
The KNT collaboration~\cite{KNT17} has kindly provided us with the values of the first few moments of the hadronic spectral function with their errors, as well as their covariance matrix. These moments were obtained using the same hadronic spectral function which results in the second number quoted in Eq.~\rf{eq:marichevend}. It  provides us with a good test of how well the approximants  that we propose work when applied to a set of hadronic moments with realistic errors. The first five moments with their errors are given in Table~\rf{table:teubner} and their correlation matrix is given in Table~\rf{table:alex} in the next section. We observe that the relative errors of the first two moments $\cM(0)$ and  $\cM(-1)$ in Table~\rf{table:teubner}  are smaller than  the relative error in the determination of the lowest order HVP contribution to $a_{\mu}^{\text{HVP}}$ in Eq.~\rf{eq:HVPexps}~\cite{KNT17}. The higher moments $\cM(-n)$ for $n=2,3,...$ have higher relative errors but they of course 
contribute  less and less to the total $a_{\mu}^{\rm HVP}$ determination.

\begin{table*}[h]
\caption[Results]{ $\cM(s)$ Moments and Errors in $10^{-3}$ units . } 
\lbl{table:teubner}
\begin{center}
\begin{tabular}{|c|c|c|} \hline \hline {\bf Moment} & {\bf Experimental Value} & {\bf Relative  Error}
\\ 
\hline \hline
$\cM(0)$~~ & $0.7176\pm 0.0026$ &  $0.36\%$
\\
$\cM(-1)$ & $0.11644\pm 0.00063$ & $0.54\%$ \\
$\cM(-2)$ & $0.03041\pm 0.00029$ & $0.95\%$ \\
$\cM(-3)$ & $0.01195\pm 0.00017$ & $1.4\%$\\
$\cM(-4)$ & $0.00625\pm 0.00011$ & $1.8\%$ \\
$\cM(-5)$ & $0.003859\pm 0.000078$ & $2.0\%$ \\
\hline\hline
\end{tabular}
\end{center}
\end{table*} 

We shall next proceed, like in the previous section, to the construction of successive MBa's of the type shown  in Eq.~\rf{eq:marichevend}  and to the evaluation of the corresponding GH-Functions  $\Pi^{\rm QCD}_{N}(Q^2)$ and $\frac{1}{\pi}\Imm\Pi_{N}(t)$. At each approximation we shall then evaluate  the corresponding   $a_{\mu}^{\rm HVP}(N)$ contribution to the muon anomlay. In the next subsection we shall only consider as input the center values of the moments in Table~\rf{table:teubner} and postpone the error analysis for later discussion in the next subsection.

\subsection{\large Successive MBa's to $\cM^{\rm QCD}(s)$, $\Pi^{\rm QCD}(Q^2)$, $\frac{1}{\pi}\Imm\Pi^{\rm QCD}(t)$ and $a_{\mu}^{\rm HVP}$.}

\noi
\subsubsection{\large\bf The $N=1$ MBa.}

\vspace*{0.25cm}
\noi
This corresponds to the MBa which one can construct when only the first moment $\cM(0)$ is known. In this case
\be\lbl{eq:mel1QCD}
\cM_{1}(s)=\frac{\alpha}{\pi}\frac{5}{3}\Gamma(1-s)\frac{\Gamma(b_{1}-1)}{\Gamma(b_{1}-s)}\,,
\ee
where the singularity at $s=1$ is the one associated to the asymptotic leading behaviour of the QCD spectral function with $u$, $d$, $s$, $c$, $b$ and $t$ quarks in Eq.~\rf{eq:pqed}.
Matching the value of $\cM_{1}(s)$ at $s=0$ with the one from the experimental determination in Table~\rf{table:teubner} fixes the $b_1$-parameter to the value:
\be
b_{1}=6.395\,.
\ee
\begin{figure}[!ht]
\begin{center}
\hspace*{-1cm}\includegraphics[width=0.40\textwidth]{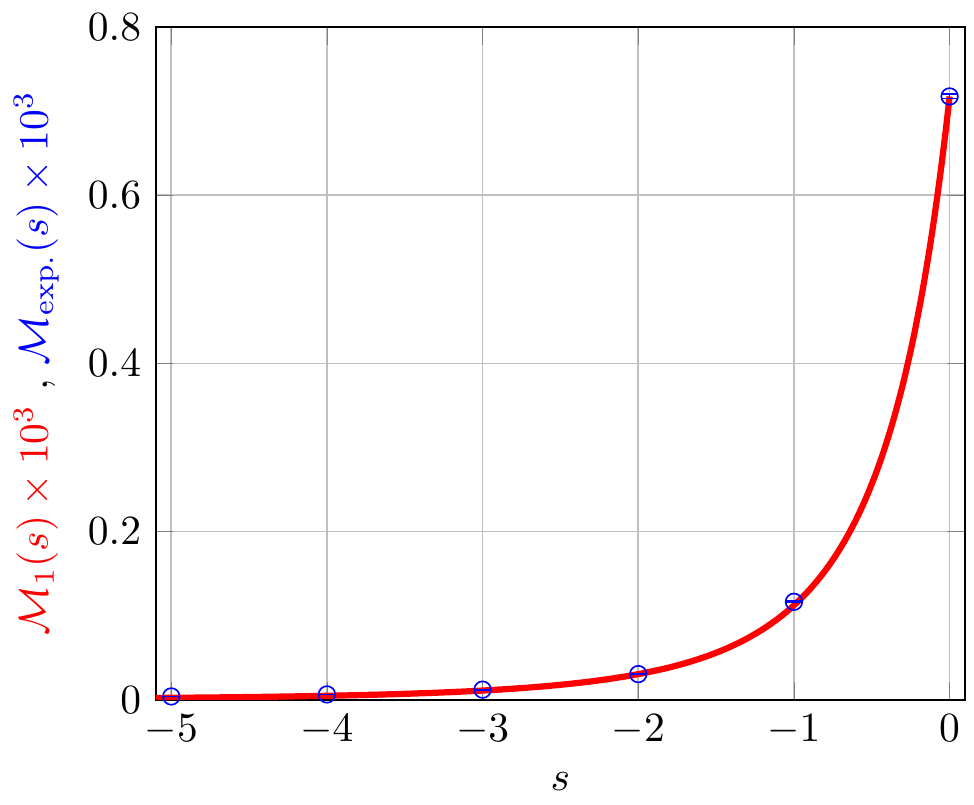} 
\bf\caption{\lbl{fig:mel1QCD}}
\vspace*{0.25cm}
{\it The red curve shows the shape of the $N=1$ MBa in Eq.~\rf{eq:mel1QCD}.\\ 
The blue circles are the experimental values in Table~\rf{table:teubner}.}
\end{center}
\end{figure}
\noi

Figure~\rf{fig:mel1QCD} shows the shape of the predicted Mellin transform. The blue points in the figure correspond  to the experimental values of the moments in Table~\rf{table:teubner} with their errors, which are too small  to be seen at the scale in the figure. The agreement, at the precision of the scale of the figure, is excellent. 

Inserting the expression of the first Mellin approximant $\cM_{1}(s)$ in the integrand at the r.h.s. of Eq.~\rf{eq:MBamu} gives the result of the first MBa to the muon anomaly:

{\setl
\bea\lbl{eq:MBaRamuFQCD}
a_{\mu}^{\rm HVP}(N=1) & = &   \left(\frac{\alpha}{\pi}\right)\sqrt{\frac{m_{\mu}^2}{t_0}}\frac{1}{2\pi }\int\limits_{-\infty}^{+\infty}d\tau \ e^{-i\tau \log\frac{t_0}{m_{\mu}^2}}\  \cF\left(\frac{1}{2}-i\tau\right)\ \cM_{N=1}\left(\frac{1}{2}-i\tau \right)\\ 
& = & 6.991\times 10^{-8}\,,
\eea}

\noi
which reproduces the central value result in Eq.~\rf{eq:HVPexps}~\cite{KNT17} surprisingly well:  to $0.8\%$.
\begin{figure}[!ht]
\begin{center}
\hspace*{-1cm}\includegraphics[width=0.40\textwidth]{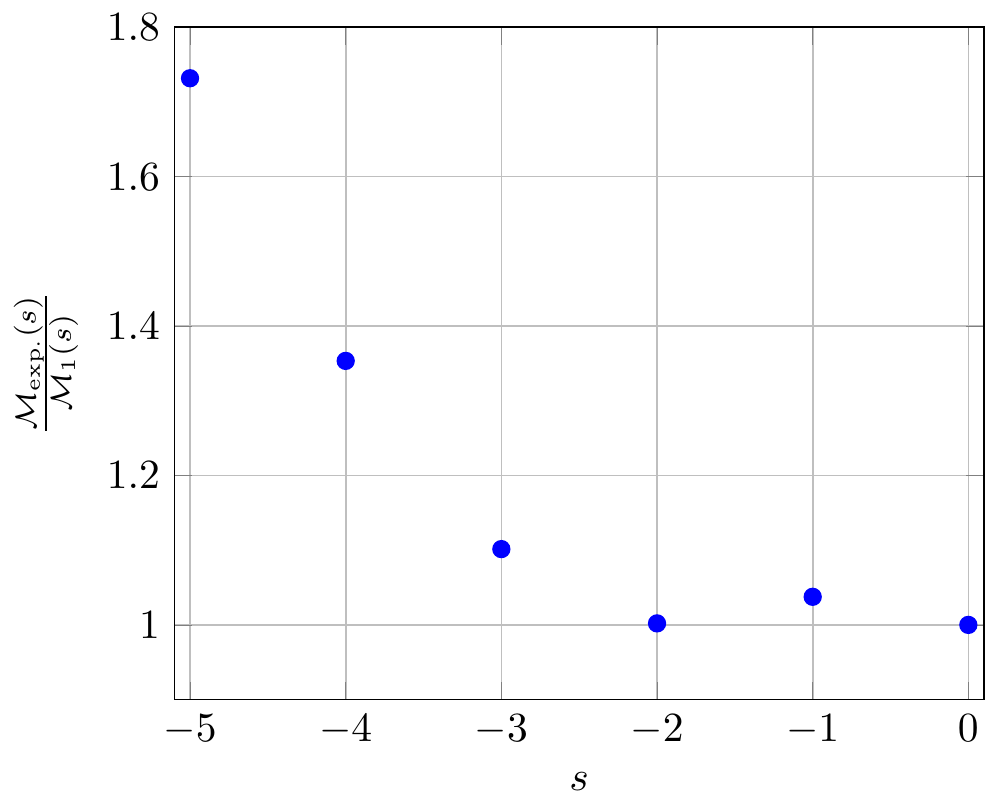} 
\bf\caption{\lbl{fig:R1E}}
\vspace*{0.25cm}
{\it Plot of the ratio of the experimental moments in Table~\rf{table:teubner} with their errors\\  to those predicted by the $N=1$ Mellin-Barnes-Approximation.}
\end{center}
\end{figure}

In order to understand why the $N=1$ MBa is already so good, let us explore more in detail the plot of $\cM_{1}(s)$ in Fig~\rf{fig:mel1QCD}. To better observe the deviations between the experimental moments and the predicted moments we plot  in Fig.~\rf{fig:R1E} their ratio as a function of $s=-n$, $n=0,1,2,\dots$.  The deviation of this ratio  from one shows the discrepancy. Notice that, here,  only the value of the $\cM(0)$ moment has been used as an input. The predicted values of $\cM(-1)$, $\cM(-2)$ and even $\cM(-3)$ turn out to be rather close to the experimental values, although already the predicted $\cM(-3)$ and certainly the predicted higher moments are not compatible with the experimental  statistical errors. Higher moments, however,  contribute less ans less to the total value of the anomaly and this is why $a_{\mu}^{\rm HVP}(N=1)$ turns out to be already such a good approximation.

Why does the $N=1$ MBa do a better job in the case of QCD  than  in the two loop QED case we discussed before? The reason for this is that in the QCD case, contrary to the QED case,  there are  resonances in the low energy region of the spectral function with mass scales which, relative to the muon mass, enhance the  contribution of the low moments, in particular $\cM(0)$. If instead of the muon anomaly we were considering the electron anomaly, the $N=1$ MBa would  already be  giving a result with an accuracy comparable to the full determination.

Although, given the result in Eq.~\rf{eq:MBaRamuFQCD} and the present accuracy from experiment, there seems to be little room for improvement, let us examine what happens when one tries the $N=2$ MBa.

\noi
\subsubsection{\large\bf The $N=2$ MBa.}

\vspace*{0.25cm}
\noi
Here the Mellin approximant has the analytic form
\be\lbl{eq:mel2QCD}
\cM_{2}(s)=\frac{\alpha}{\pi}\frac{5}{3}\frac{\Gamma(1-s)}{\Gamma(2-s)}
\frac{\Gamma(a_1 -s)}{\Gamma(a_1 -1)}\frac{\Gamma(b_1 -1)}{\Gamma(b_1 -s)}\,,
\ee
and the parameters $a_1$ and $b_1$ are fixed by the matching equations:
\be\lbl{eq:a1b1}
\cM_{2}(0)=\cM(0)\quad\annd\quad \cM_{2}(-1)=\cM(-1) \,,
\ee
with  $\cM(0)$ and $\cM(-1)$ given in Table~\rf{table:teubner}. This results in the values:
\be\lbl{eq:HVPab}
a_1 =1.900\quad\annd\quad b_1 = 5.855\,.
\ee
The shape of the $\cM_{2}(s)$ Mellin transform turns out to be  rather similar to the $\cM_{1}(s)$ one in Fig.~\rf{fig:R1E}. In order to appreciate the differences between the $N=1$ and  $N=2$ MBa's, we compare in Fig.~\rf{fig:R12E} the ratios of the experimental moments to those of the $\cM_{2}(s)$ prediction (the red dots) and to those of the $\cM_{1}(s)$ prediction (the blue dots). The overall shape of the red dots is clearly better because they are nearer to one.

\begin{figure}[!ht]
\begin{center}
\hspace*{-1cm}\includegraphics[width=0.50\textwidth]{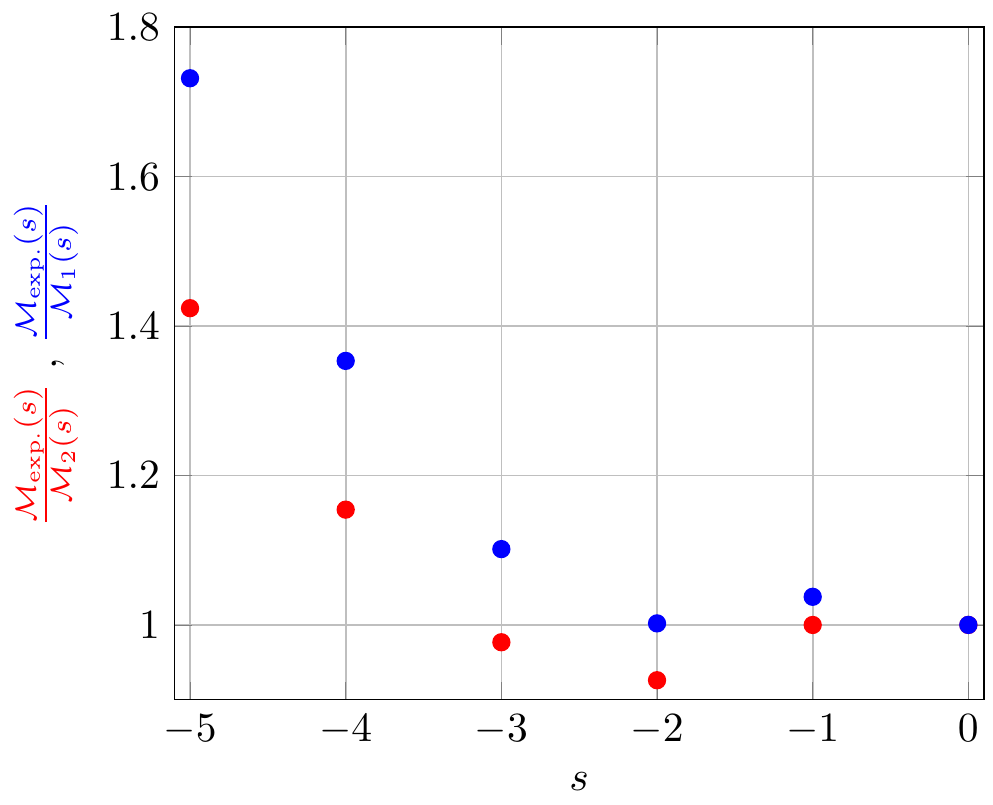} 
\bf\caption{\lbl{fig:R12E}}
\vspace*{0.25cm}
{\it Plot of the ratio of the experimental moments in Table~\rf{table:teubner} with their errors\\  to those predicted by the $N=2$ MBa in red and the $N=1$ MBa in blue.}
\end{center}
\end{figure}

With the expression of the second Mellin approximant $\cM_{2}(s)$ inserted in the integrand at the r.h.s. of Eq.~\rf{eq:MBamu} we get as  a result of the $N=2$ MBa to the muon anomaly:

{\setl
\bea\lbl{eq:MBaRamuFQCD2}
a_{\mu}^{\rm HVP}(N=2) & = &   \left(\frac{\alpha}{\pi}\right)\sqrt{\frac{m_{\mu}^2}{t_0}}\frac{1}{2\pi }\int\limits_{-\infty}^{+\infty}d\tau \ \underbrace{e^{-i\tau \log\frac{t_0}{m_{\mu}^2}}\  \cF\left(\frac{1}{2}-i\tau\right)\ \cM_{N=2}\left(\frac{1}{2}-i\tau \right)}_{\cR(\tau)}\\ 
 & = & 6.970\times 10^{-8}\,,
\eea}

\noi
which reproduces the central value result in Eq.~\rf{eq:HVPexps}~\cite{KNT17} at the $0.5\%$ level, i.e. an improvement by a factor of 1.6 with respect to the $N=1$ case. Figure~\rf{fig:integrand2} shows the shape of the integrand $\cR(\tau)$ in Eq.~\rf{eq:MBaRamuFQCD2} which, as expected,  has a rapid decrease as $\vert\tau\vert\gtrsim 1$.

\begin{figure}[!ht]
\begin{center}
\hspace*{-1cm}\includegraphics[width=0.50\textwidth]{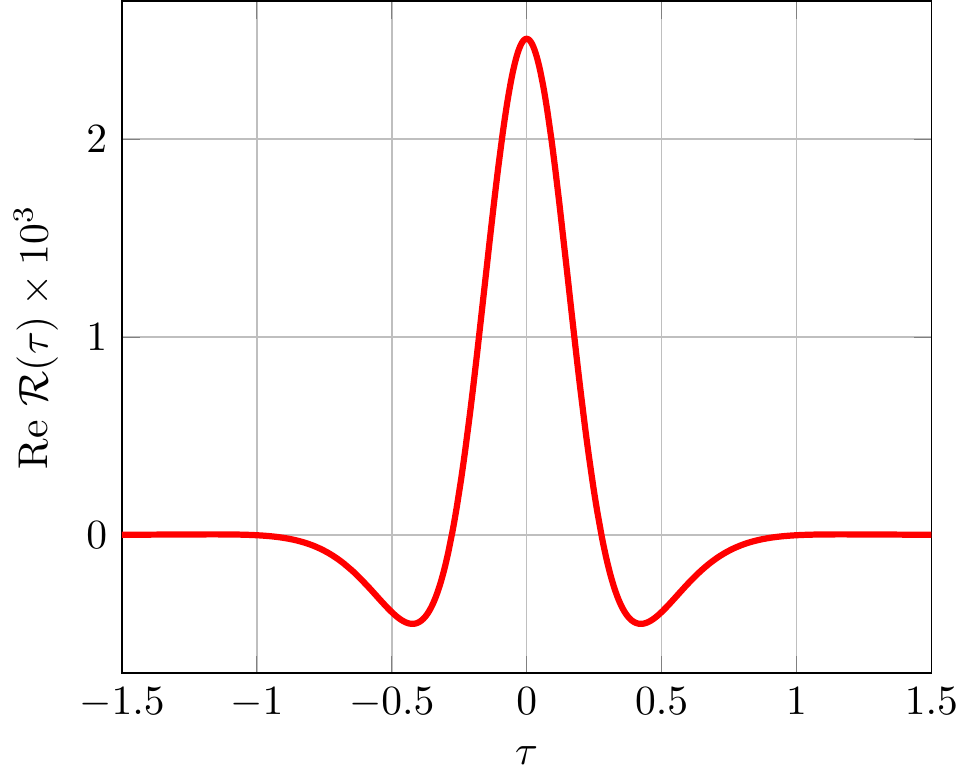} 
\bf\caption{\lbl{fig:integrand2}}
\vspace*{0.25cm}
{\it Plot of the integrand in Eq.~\rf{eq:MBaRamuFQCD2} as a function of $\tau$.}
\end{center}
\end{figure}

As discussed in the previous section, the MBa technique allows to reconstruct as well $\Pi_{N}(Q^2)$ approximants of the HVP self energy in terms of GH-functions. The corresponding $N=2$ approximant is ($z=\frac{Q^2}{t_0}$):
\be\lbl{eq:meijerN2QCD}
\Pi_{N=2}^{\rm QCD}(Q^2) =  \left(\frac{\alpha}{\pi} \right) \ (-z)\frac{5}{3}\frac{a_1 -1}{b_1 -1}\ _{3}{F}_{2}\left(\left. \begin{array}{ccc} 1 & 1 & a_1 \\ ~ & 2 & b_1 \end{array}\right\vert {-z}\right)\,,
\ee
with $a_1$ and $b_1$ given in Eq.~\rf{eq:HVPab}. 
The shape of the function $\Pi_{N=2}^{\rm QCD}(Q^2)$ is shown in Fig.~\rf{fig:PI2QCD}.

\begin{figure}[!ht]
\begin{center}
\hspace*{-1cm}\includegraphics[width=0.50\textwidth]{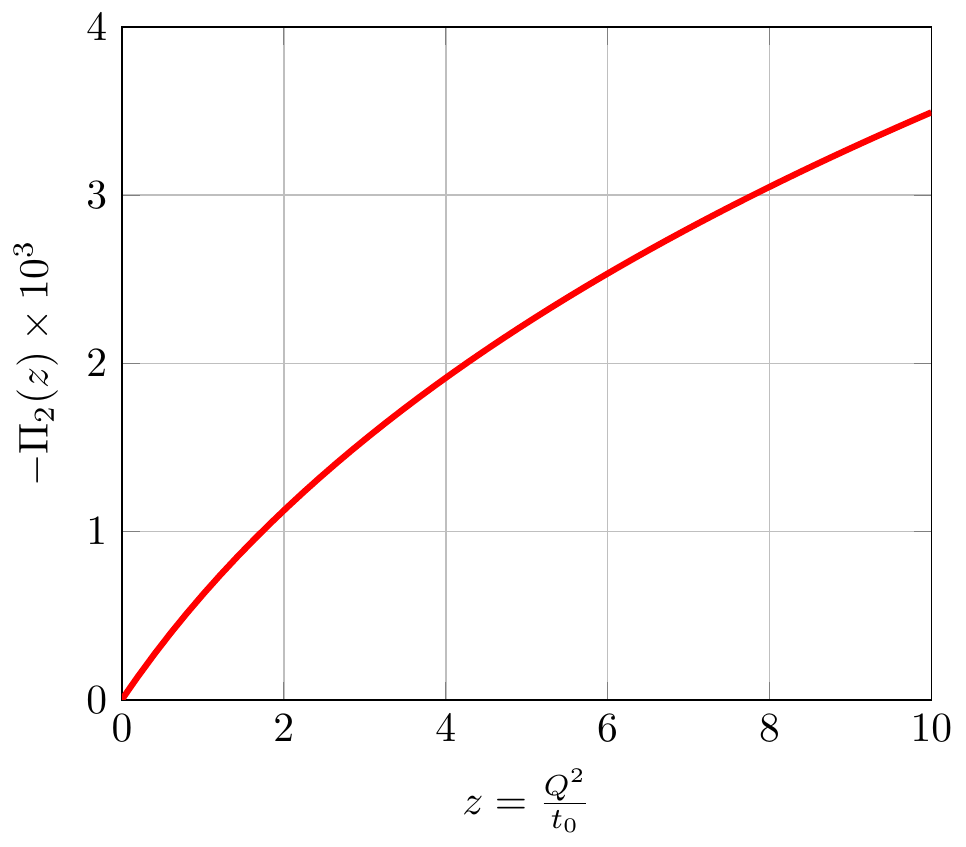} 
\bf\caption{\lbl{fig:PI2QCD}}
\vspace*{0.25cm}
{\it Shape of the function $\Pi_{N=2}^{\rm QCD}(Q^2)$ in Eq.~\rf{eq:meijerN2QCD} as a function of $z=\frac{Q^2}{t_0}$.}
\end{center}
\end{figure}
\begin{figure}[!ht]
\begin{center}
\hspace*{-1cm}\includegraphics[width=0.50\textwidth]{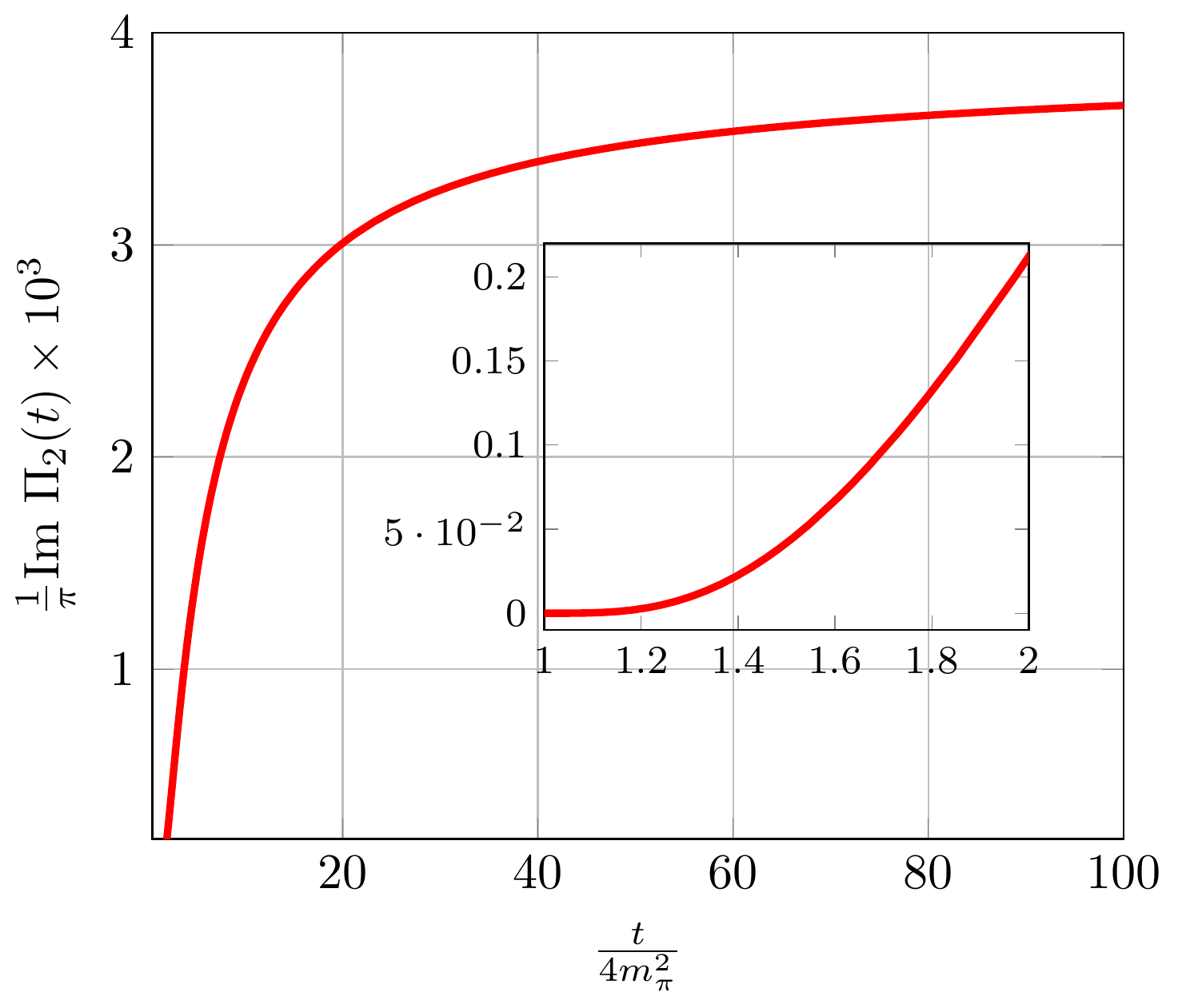}
\bf\caption{\lbl{fig:spec2QCD}}
\vspace*{0.25cm}
{\it Plots of the $N=2$ MBa Spectral Function. }
\end{center}
\end{figure}

Plots of the spectral function associated to the $N=2$ MBa are also shown in Figs.\rf{fig:spec2QCD}. Although, asymptotically, the $N=2$ MBa spectral function approaches the pQCD value it can  only be considered a smooth interpolation of the physical spectral function which, as we know, has a lot of local structure.  This interpolation, however,  when inserted in the r.h.s. of Eq.~\rf{eq:str} reproduces the determination of the anomaly using the experimental spectral function at the { $0.5\%$} level already mentioned. It is in this sense that it is a good interpolation.

We shall next explore what happens when one tries to improve the $N=2$ MBa with higher approximants and further input from the experimental values of higher moments.

\noi
\subsubsection{\large\bf The $N=3$ MBa.}

\vspace*{0.25cm}
\noi
The corresponding Mellin approximant which generalizes the one in Eq.~\rf{eq:mel1QCD} has the analytic form
\be\lbl{eq:mel3QCD}
\cM_{3}(s)=\frac{\alpha}{\pi}\frac{5}{3}\Gamma(1-s)\frac{\Gamma(b_{1}-1)}{\Gamma(b_{1}-s)}\frac{\Gamma(a_{1}-s)}{\Gamma(a_{1}-1)}\frac{\Gamma(b_{2}-1)}{\Gamma(b_{2}-s)}\,,
\ee
with the parameters $a_1$, $b_1$ and $b_2$ solutions of the matching equations
\be\lbl{eq:match3}
\cM_{3}(0)=\cM(0)\,,\quad\cM_{3}(-1)=\cM(-1)\quad\annd\quad \cM_{3}(-2)=\cM(-2)\,.
\ee
In this case one finds a ``possible solution'' where
\be\lbl{eq:nogood}
a_1 =-0.362\,,\quad b_1 =6.462\,,\quad\ b_2 =-0.346\,,
\ee
and the equivalent one with $b_{1}\rightleftharpoons b_{2}$. These ``solutions'', however,  are not acceptable because they generate a pole at $s=a_1$ which is inside of the fundamental strip in contradiction with first principles, as discussed in Section III.3. Nevertheless,  the negative  numerical values of $a_1$ and $b_2$ are in fact rather close to each other. Had they been exactly the same,  there would have  been a cancellation between $\Gamma(a_1 -s)$ and $\Gamma(b_2 -s)$ in Eq.~\rf{eq:mel3QCD} indicating that it is not possible to improve beyond $N=2$ with a single Marichev-like function. The situation here is rather similar to the one encountered earlier when considering the $N=6$ MBa in the QED example.     

The fact that in QCD the simple Marichev-like approximants   fail to find physical solutions already at the $N=3$ level is perhaps not so surprising. One does not expect,  beyond a certain level of accuracy,  to be able to approximate $\Pi^{\rm QCD}(Q^2)$ at all $Q^2$ values with just one GH-function. One may, however, ask: is it possible  to find  generalizations of the simple Marichev-like MBa's which, when using  more than the first two moments in Table~\rf{table:teubner} as an input, provide acceptable solutions to compare with $a_{\mu}^{\rm HVP}$ in Eq.~\rf{eq:HVPexps}~\cite{KNT17}? As already mentioned at the end of Section IV there is a positive answer to that. It consists in using  standard superpositions of  Mellin approximants of the type indicated in Eq.~\rf{eq:marichevend}. This, in turn, implies specific superpositions of GH-Functions which approximate  the self-energy $\Pi^{\rm QCD}(Q^2)$ in the Euclidean,  and hence $a_{\mu}^{\rm HVP}$.

\noi
\subsubsection{\large\bf The $N=(2)+(1)$ MBa.}

\vspace*{0.25cm}
\noi
The simplest superposition which gives acceptable solutions to the matching equations, when one knows three moments in the HVP case, consists of the sum of one $N=2$ MBa and one $N=1$ MBa:
\be\lbl{eq:mel21QCD}
\cM_{2+1}(s)=\frac{\alpha}{\pi}\frac{5}{3}\frac{1}{2}\left\{\frac{1}{1-s}\frac{\Gamma(a_1 -s)}{\Gamma(a_1 -1)}\frac{\Gamma(b_1-1)}{\Gamma(b_1-s)}+\Gamma(1-s)\frac{\Gamma(b_{2}-1)}{\Gamma(b_2-s)}\right\}\,,
\ee
with the overall factor 1/2  fixes the correct pQCD residue at $s=1$,
and the parameters $a_1$, $b_1$ and $b_2$ are solutions of the matching equations:
\be\lbl{eq:2+1}
\cM_{2+1}(0)=\cM(0)\,,\quad\cM_{2+1}(-1)=\cM(-1)\quad\annd\quad\cM_{2+1}(-2)=\cM(-2)\,.
\ee
There is  only one acceptable solution to these equations with the values:
\be
a_1=5.2668,\quad b_1=14.514\,,\quad\annd\quad b_2=19.177\,.
\ee
With $\cM_{2+1}(s)$ inserted in the integrand at the r.h.s. of Eq.~\rf{eq:MBamu} we get as  a result for the muon anomaly:
\be
a_{\mu}^{\rm HVP}(N=2+1)=6.957\times 10^{-8}
\ee
which reproduces the central value result in Eq.~\rf{eq:HVPexps}~\cite{KNT17} at the $0.4\%$ level, and is an improvement  with respect to the previous $N=2$ case.
\begin{figure}[!ht]
\begin{center}
\hspace*{-1cm}\includegraphics[width=0.50\textwidth]{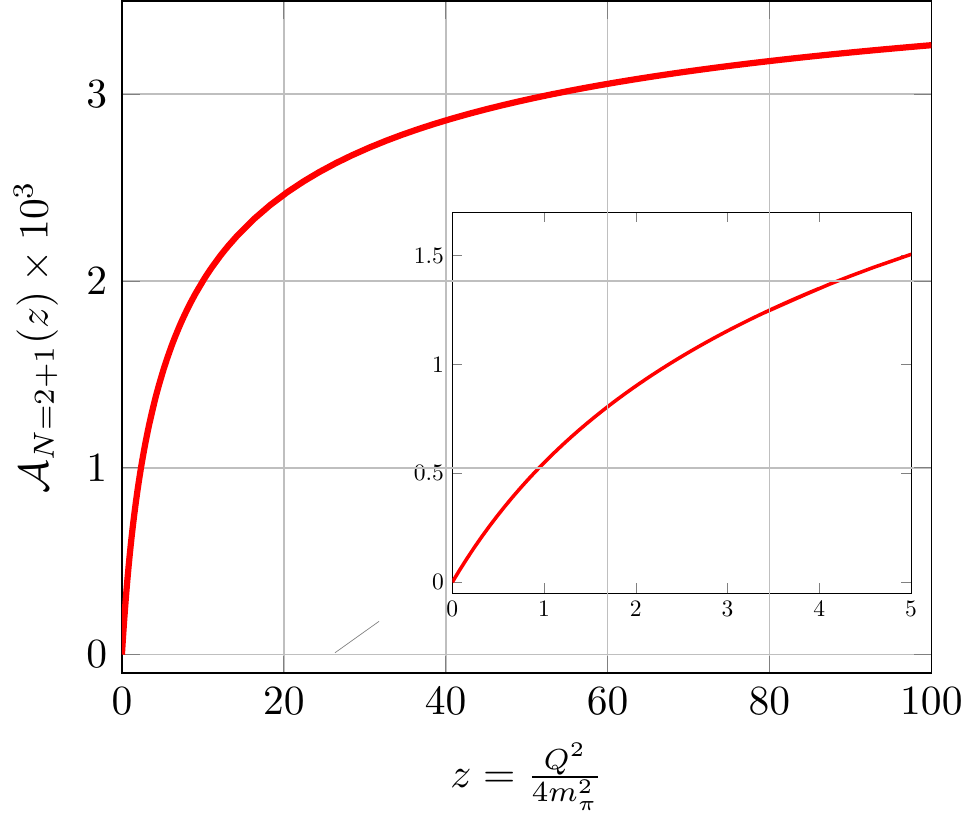}
\bf\caption{\lbl{fig:adler2+1}}
\vspace*{0.25cm}
{\it Plots of the $N=2+1$ Adler Function versus $z=\frac{Q^2}{t_0}$. }
\end{center}
\end{figure}

The corresponding sum of HG-Functions to the $\cM_{2+1}(s)$ MBa in Eq.~\rf{eq:2+1} which results as an approximation to the HVP self-energy is now

{\setl
\bea\lbl{eq:2+1-Pi}
\Pi_{N=2+1}^{\rm QCD}(Q^2) &  = &   \left(\frac{\alpha}{\pi} \right)\   (-z)\frac{5}{3}\frac{1}{2}  \left\{\frac{a_1 -1}{b_1 -1}\ _{3}{F}_{2}\left(\left. \begin{array}{ccc} 1 & 1 & a_1 \\ ~ & 2 & b_1 \end{array}\right\vert {-z}\right)\right.\nn \\
 &  & \hspace*{2.5cm} +\left.\frac{1}{b_2 -1}\ _{2}{F}_{1}\left(\left. \begin{array}{cc} 1 & 1 \\ ~ & b_2\end{array}\right\vert {-z}\right) \right\}\,,
\eea}

\noi
and the corresponding approximation to the Adler function is

{\setl
\bea\lbl{eq:2+1Ad}
\cA_{N=2+1}^{\rm QCD}(Q^2) &  = &   \left(\frac{\alpha}{\pi} \right)\   z\frac{5}{3}\frac{1}{2}  \left\{\frac{a_1 -1}{b_1 -1}\ _{3}{F}_{2}\left(\left. \begin{array}{ccc} 2 & 1 & a_1 \\ ~ & 2 & b_1 \end{array}\right\vert {-z}\right)\right.\nn \\
 &  & \hspace*{2cm} +\left.\frac{1}{b_2 -1}\ _{2}{F}_{1}\left(\left. \begin{array}{cc} 2 & 1 \\ ~ & b_2\end{array}\right\vert {-z}\right) \right\}\,.
\eea}

\noi
The shape of this Adler function is shown in Fig.~\rf{fig:adler2+1}.

\noi
\subsubsection{\large\bf The $N=(2)+(1)+(1)$ MBa.}

\vspace*{0.25cm}
\noi
With the first four moments of HVP as an input, there is
a new superposition of MBa's which gives an acceptable solution to the matching equations. It is the following linear combination of a $N=2$ MBa and two $N=1$ MBa's:
\be\lbl{eq:2+1+1}
\cM_{2+1+1}(s)=\frac{\alpha}{\pi}\frac{5}{3}\left\{\frac{1}{1-s}\frac{\Gamma(a_1 -s)}{\Gamma(a_1 -1)}\frac{\Gamma(b_1 -1)}{\Gamma(b_1 -s)}+ \Gamma(2-s)\frac{\Gamma(b_2 -1)}{\Gamma(b_2 -s)}+ 
\Gamma(2-s)\frac{\Gamma(b_3 -1)}{\Gamma(b_3 -s)}\right\}\,.
\ee

\begin{figure}[!ht]
\begin{center}
\hspace*{-1cm}\includegraphics[width=0.50\textwidth]{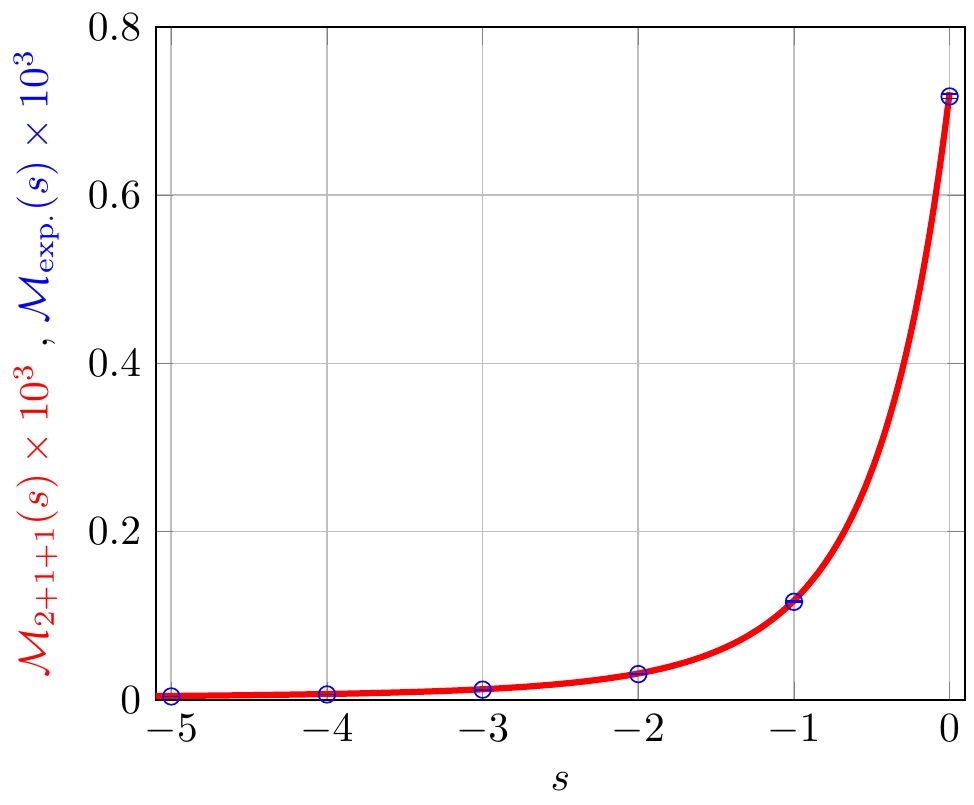} 
\bf\caption{\lbl{fig:2+1+1}}
\vspace*{0.25cm}
{\it The red curve is the shape of  $\cM_{2+1+1}$ in Eq.~\rf{eq:2+1+1} for $-5\le s\le  0$.\\ The dots are the experimental values of the moments.}
\end{center}
\end{figure}

\noi
The matching equations:

{\setl
\bea
\lefteqn{\cM_{2+1+1}(0)=\cM(0)\,,\quad\cM_{2+1+1}(-1)=\cM(-1)\,,} \nn\\
& & \cM_{2+1+1}(-2)=\cM(-2)\,,\quad\annd\quad\cM_{2+1}(-3)=\cM(-3)\,,
\eea}

\noi
give  an acceptable solution with values:
\be
a_1 = 1.0180\,,\quad b_1=1.7495\,,
\ee
and two complex conjugate values for $b_2$ and $b_3$, or equivalently $b_{2}\rightleftharpoons b_{3}$:
\be
b_2 =12.822 + i~2.6069\,,\quad b_3 =12.822 - i~2.6069\,,
\ee
which gives  a total real contribution to the sum of the two $N=1$ terms in Eq.~\rf{eq:2+1+1}.

The expression of the $N=2+1+1$ Mellin approximant $\cM_{2+1+1}(s)$ inserted in the integrand at the r.h.s. of Eq.~\rf{eq:MBamu} results in a value  for the muon anomaly:
\be
a_{\mu}^{\rm HVP}(N=2+1+1)=6.932\times 10^{-8}\,,
\ee
which almost exactly reproduces the central value result in Eq.~\rf{eq:HVPexps}~\cite{KNT17}, and represents a net improvement  with respect to the previous $N=2+1$ approximation.

The shape of the Mellin transform $\cM_{2+1+1}(s)$ is shown in Fig.~\rf{fig:2+1+1} together with the experimental values of the first five moments. Figure~\rf{fig:ratio2+1+1} shows the ratio of the experimental values of the first five moments to the values predicted by $\cM_{2+1+1}$ in Eq.~\rf{eq:2+1+1}.

\begin{figure}[!ht]
\begin{center}
\hspace*{-1cm}\includegraphics[width=0.50\textwidth]{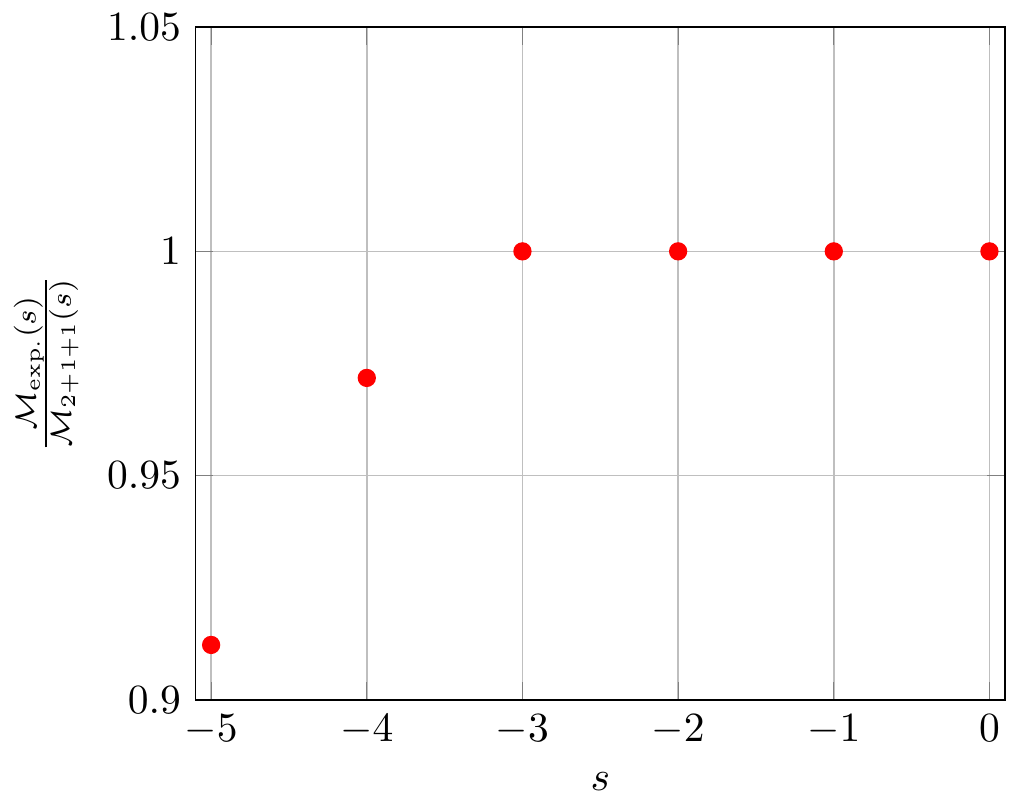} 
\bf\caption{\lbl{fig:ratio2+1+1}}
{\it Plot of the ratio of the experimental moments in Table~\rf{table:teubner} to those of the $N=2+1+1$ MBa.\\
Notice the difference of scale in the vertical axis, as compared to the one in Fig.~17.}
\end{center}
\end{figure}
\vspace*{-0.25cm}
The Adler function associated to $\cM_{2+1+1}(s)$ in Eq.~\rf{eq:2+1+1} is the sum of three GH-Functions:

{\setl
\bea\lbl{eq:2+1+1Ad}
\cA_{N=2+1+1}^{\rm QCD}(Q^2) &  = &   \left(\frac{\alpha}{\pi} \right)\   z\frac{5}{3}  \left\{\frac{a_1 -1}{b_1 -1}\ _{3}{F}_{2}\left(\left. \begin{array}{ccc} 2 & 1 & a_1 \\ ~ & 2 & b_1 \end{array}\right\vert {-z}\right)\right.\nn \\
 &  &  +\left.\frac{1}{b_2 -1}\ _{2}{F}_{1}\left(\left. \begin{array}{cc} 2 & 2 \\ ~ & b_2\end{array}\right\vert {-z}\right) + \frac{1}{b_3 -1} \ _{2}{F}_{1}\left(\left. \begin{array}{cc} 2 & 2 \\ ~ & b_3\end{array}\right\vert {-z}\right) \right\}\,,
\eea}
\begin{figure}[!ht]
\begin{center}
\hspace*{-1cm}\includegraphics[width=0.50\textwidth]{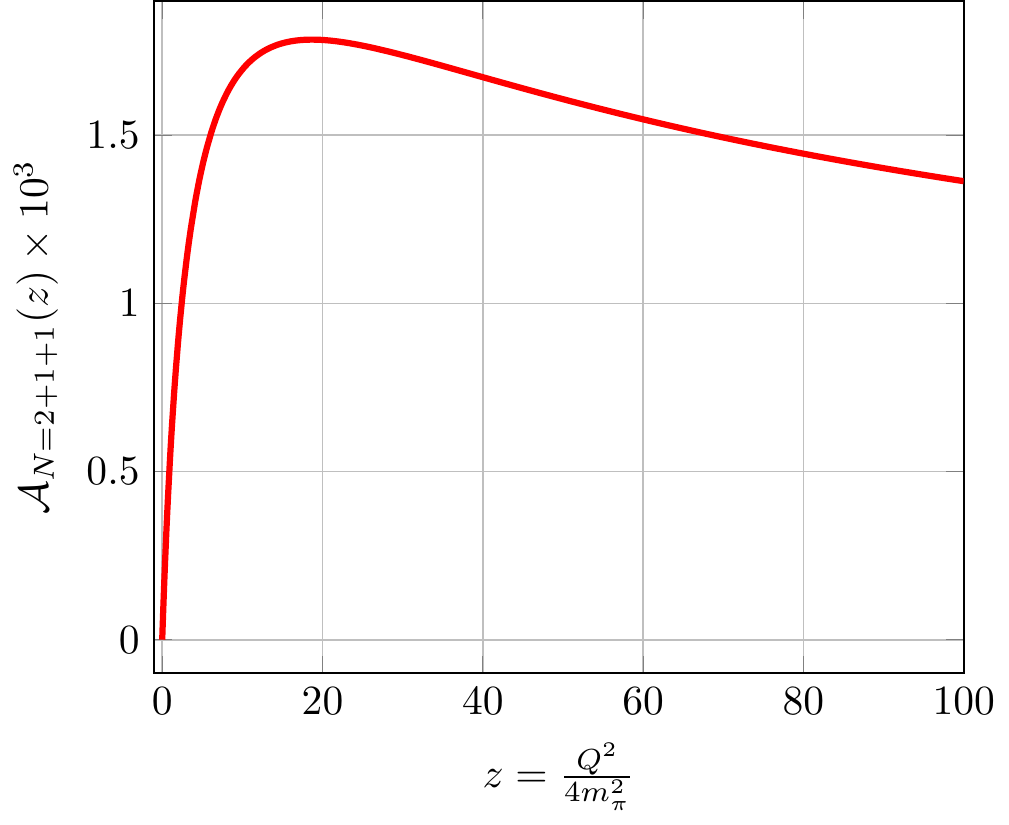} 
\bf\caption{\lbl{fig:ad2+1+1}}
\vspace*{0.25cm}
{\it Plot of the Adler function in Eq.~\rf{eq:2+1+1Ad}.}
\end{center}
\end{figure}

\noi
and its shape is shown in Fig.~\rf{fig:ad2+1+1}.

\noi 

Plots of the spectral function  corresponding to the $N=2+1+1$
 MBa are also shown in Fig.~\rf{fig:spect211}. {  The plots already exhibit underlying features of the hadronic structure.}

\begin{figure}[!ht]
\begin{center}
\hspace*{-1cm}\includegraphics[width=0.40\textwidth]{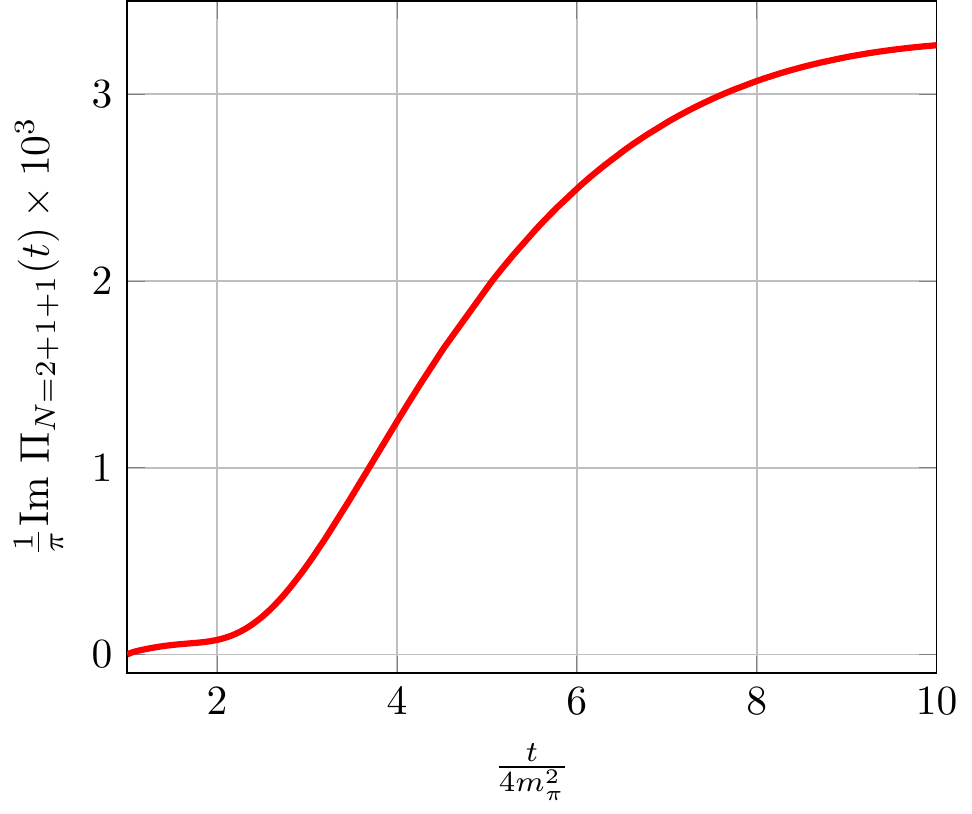} \includegraphics[width=0.40\textwidth]{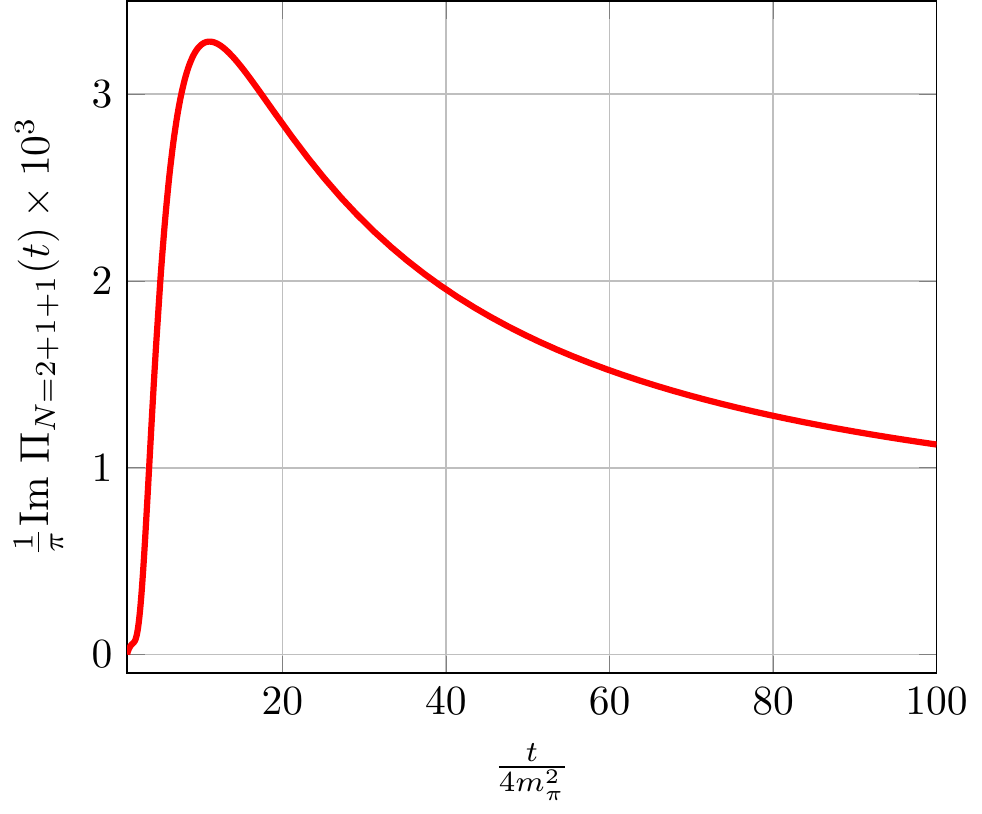} 
\bf\caption{\lbl{fig:spect211}}
\vspace*{0.25cm}
{\it Plots of the $N=2+1+1$ Spectral Function. }
\end{center}
\end{figure}

\noi
\subsection{\large Uncertainties of the Successive MBa's to $a_{\mu}^{\rm HVP}$.}

\vspace*{0.25cm}
\noi

We shall finally examine the sensitivity of the results obtained for the $a_{\mu}^{\rm HVP}(N)$ to small variations in the input parameters $a_k$ and $b_k$ of the successive $\cM_{N}(s)$, as well as to the choice of the $N$-approximant itself. The errors in the experimental determination of the moments $\cM(-n)$ have been tabulated in Table~\rf{table:teubner} and their correlation matrix is given in Table~\rf{table:alex}. One can see that the values of these moments  are highly correlated, reflecting the fact that they all have been extracted from different integrals of the same input data on the spectral function.

The statistical part of the analysis is standard. We first construct the covariance matrix $C_{ij}$ of the first $N$  moments obtained from experiment $\cM(1-i)\,,i=1,\dots,N$:
\be
C_{ij}=\rho_{ij}\sigma_i\sigma_j\,,\quad\text{with}\quad \rho_{ii}=1\,,\ \  -1<\rho_{i,j}<+1\quad\annd\quad i,j=1,\dots,N\,,
\ee
where $\rho_{ij}$ is the correlation coefficient between the moment $\#i$ and the moment $\#j$, each with Gaussian uncertainty $\sigma_{i}$ and $\sigma_{j}$ . Then we define a $\chi^2$ function associated to a given Mellin-Barnes approximant $\cM_{N}(s)$, which depends on a  set of parameters $(a_k\,,b_k)$:
\begin{equation}\label{eq:chi2}
\chi^2 = \sum_{i,j=1}^{N} \left[\cM_{N}(1-i)-\cM(1-i)\right] C^{-1}_{ij} \left[\cM_{N}(1-j)-\cM(1-j)\right]\,.
\end{equation}

\begin{table*}[h]
\caption[Results]{\it Correlation Matrix of the Moments $\cM(0),\ldots,\cM(-5)$ in Table~\rf{table:teubner}  } 
\lbl{table:alex}
\begin{displaymath}
\left(\begin{array}{cccccc}
1 & 0.83 & 0.62 & 0.50 & 0.42 & 0.37 \\
& 1 &  0.93 & 0.84 & 0.77 & 0.70 \\
&& 1 & 0.98 & 0.93 & 0.88 \\
&&& 1 & 0.987 & 0.96 \\
&&&& 1 &  0.991 \\
&&&&& 1
\end{array}\right)\,.
\end{displaymath}
\end{table*}

\noi
and minimize this $\chi^2$  with respect to the set of parameters $(a_k\,,b_k)$. The errors are sufficiently small to ensure that a point-like estimate is an excellent approximation, and we obtain the covariance matrix in the $(a_k,b_k)$ parameter space from the Hessian matrix of the $\chi^2$ function computed at its minimum. Using linear error propagation we can then calculate the statistical uncertainty on $a_{\mu}^{\rm HVP}$, as reported in the third column of Table~\rf{table:uncertainties}. 
The fact that all the approximants have a similar uncertainty that coincides with the one of the complete  evaluation of $a_{\mu}^{\rm HVP}$~\cite{KNT17}  is a sign that the statistical information is saturated by all our MBa's.

\begin{table*}[h]
\caption{  Numerical results on the determination of $a_{\mu}^{\rm HVP}$ ($10^{-8}$ units), for each considered MBa. } 
\lbl{table:uncertainties}
\begin{center}
\begin{tabular}{|c|c|c|} \hline \hline {\bf MBa Ansatz} & {\bf Central Value} & {\bf Stat. Uncertainty} 
\\ 
\hline \hline
Eq.~\rf{eq:mel1QCD} ($N=1$) & 6.991 & 0.023  \\
Eq.~\rf{eq:mel2QCD} ($N=2$) & 6.970 & 0.024  \\
Eq.~\rf{eq:mel21QCD} ($N=(2)+(1)$) & 6.957 & 0.025   \\
Eq.~\rf{eq:2+1+1} ($N=(2)+(1)+(1)$) & 6.932 & 0.025  \\
\hline\hline
\end{tabular}
\end{center}
\end{table*} 

\begin{figure}[!ht]
\begin{center}
\hspace*{-1cm}\includegraphics[width=0.75\textwidth]{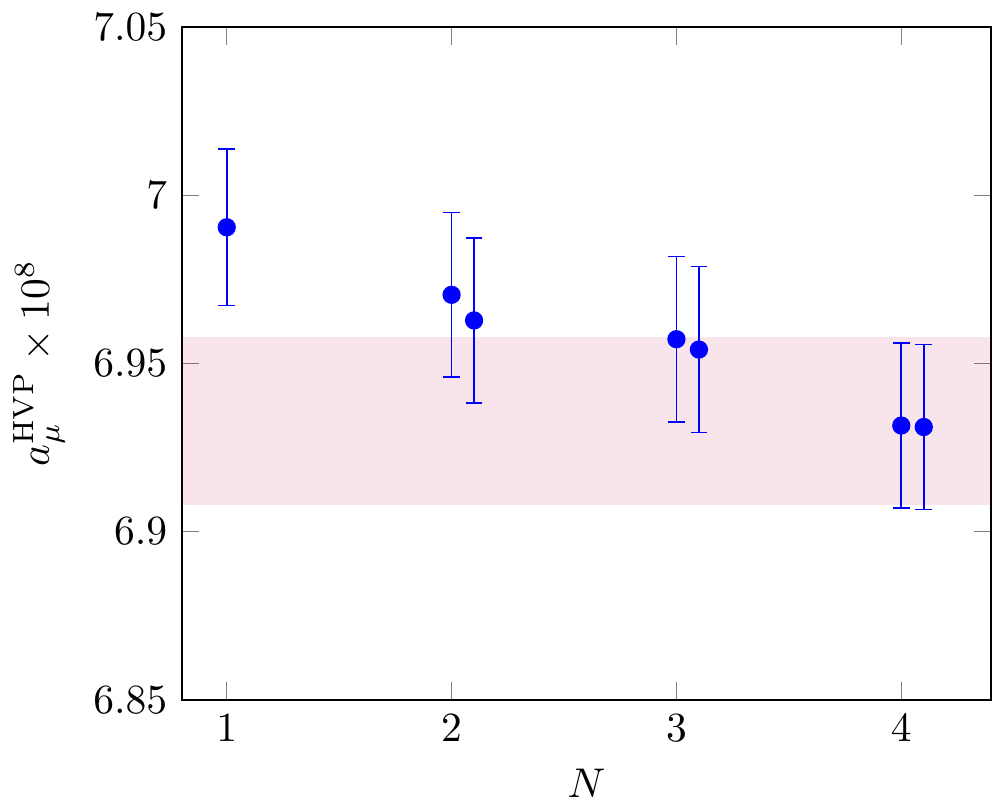} 
\bf\caption{\lbl{fig:amuN}} 
\vspace*{0.25cm}
{\it Results for $a_{\mu}^{\rm HVP}$  as a function of the number of input moments $N$. The blue points correspond to alternative choices of  MBa's (two choices for $N=2,3,4$) with their statistical uncertainty.\\{ The pink band is the full experimental result of ref.~\cite{KNT17}.}}
\end{center}
\end{figure}

Our results would not be complete without a study of the systematic shift associated to the successive MBa's which interpolate the values of the experimental moments and reconstruct the full Mellin functions. With this aim, in addition to the MBa's discussed in detail in the previous section, we have also tested alternative parameterizations for $N=2,3,4$ which are obtained by changing the location of the poles in the superposition terms ( \textit{e.g.} $\Gamma(2-s)$ instead of $\Gamma(1-s)$ in Eq.~\rf{eq:mel21QCD}). These alternative MBa's have also valid solutions for the corresponding $(a_k\,, b_k)$ parameters and, therefore, can also be considered as good  alternative choices. The results of all the evaluations of $a_{\mu}^{\rm HVP}$ which we have made are plotted in Fig.~\rf{fig:amuN},  as a function of the number of input moments $N$. 
 We observe that the successive results converge towards the  experimental value in Eq.~\rf{eq:HVPexps}.

\section{\Large Conclusions and Outlook}
\setcounter{equation}{0}
\def\theequation{\arabic{section}.\arabic{equation}}

\noi
Equation \rf{eq:momeucl} shows that moments of the hadronic spectral function are equivalent to derivatives of the hadronic self-energy function $\Pi(Q^2)$ at $Q^2 =0$. The latter are accessible to LQCD simulations as well as to eventual dedicated experiments. We have shown how, from an accurate determination of the first few moments,  one could reach an evaluation of the HVP contribution to the muon anomaly with a competitive precision, or even higher, than the present experimental determinations.

 The method that we propose uses  a new technique of Mellin-Barnes approximants which has been explained and justified in detail in the text. Essentially it is based on generic QCD properties which fix the class of Mellin transforms $\cM(s)$ of the spectral function that one can use as successive approximants. The muon anomaly $a_{\mu}^{\rm HVP}$, in terms of these $\cM(s)$-functions, is given by the Fourier transform in Eq.~\rf{eq:MBamuF}. The corresponding approximations to the hadronic self-energy function $\Pi(Q^2)$ are well defined Generalized Hypergeometric Functions which we have given explicitly and the approximations to the spectral function are also given in terms of  Meijer's G-Functions. This offers the possibility of applying the same techniques developped here  to the case where the information from LQCD, or from experiment, is given in terms of determinations of the self-energy function $\Pi(Q^2)$ at fixed Euclidean $Q^2$-values, as e.g. in ref.~\cite{Lellouch17}. { We plan to discuss this in 
the near future.}

  We have illustrated the practical application of the method   with the example of the QED contribution to the muon anomaly from the vacuum polarization Feynman diagrams in Fig.~\rf{fig:QED4}. We have also discussed the case where one uses as an input the experimental values of the first moments provided to us by the collaboration of ref.~\cite{KNT17}. We find that, in this case, our approach reproduces very well their complete phenomenological analysis.

\vspace*{0.5cm}

\begin{center}
{\Large\bf Acknowledgments}

\end{center}

\vspace*{0.25cm}

\noi
We are very grateful to Thomas Teubner and to Alex Keshavarzi for providing us with the experimental values of the first few moments and the error correlations { of their update}.  We also thank Laurent Lellouch and Ruth Van de Water for their interest and informative discussions, and Alex Keshavarzi, Ruth Van de Water and the referee for a careful reading of the manuscript. D.G. thanks M.~Knecht and CPT for their hospitality during the beginning of this work.

The work of J.C. and E.deR. has been carried out thanks to the support of the OCEVU Labex (ANR-11-LABX-0060) and the A*MIDEX project (ANR-11-IDEX-0001-02) funded by the "Investissements d'Avenir" French government program managed by the ANR.

\vspace*{1cm}

\begin{appendix}
\renewcommand{\thesection}{\normalsize \Alph {section}}

\begin{center}
{\bf\normalsize APPENDIX}
\end{center}

\vspace*{0.5cm}
\noi In this appendix we discuss various technical details which appear in the main text 

\section{\normalsize The Basic Mellin-Barnes Identity}
\setcounter{equation}{0}
\def\theequation{\Alph{section}.\arabic{equation}}

\noi
The identity in Eq.~\rf{eq:MBaid} is a particular case of the identity ($N=1,2,3,\dots$):
\be\lbl{eq:MBaidn}
\frac{1}{(1+A)^{N}}=\frac{1}{2\pi i}\int\limits_{c_s-i\infty}^{c_s+i\infty}ds \left(A \right)^{-s} \frac{\Gamma(s)\Gamma(N-s)}{\Gamma(N)}\,.
\ee
We shall first show how performing the integral in the r.h.s. for $N=1$ reproduces the l.h.s. For that we make a choice of $s$ with $\Ree(s) \in ]0,1[$, e.g. $s=\frac{1}{2}+i\tau$. Then

{\setl
\bea
\lefteqn{\frac{1}{2\pi i}\int\limits_{c_s-i\infty}^{c_s+i\infty}ds \left(A \right)^{-s} \Gamma(s)\Gamma(1-s)} \nn\\
& & = \frac{1}{\sqrt{A}}\frac{1}{2\pi}\int_{-\infty}^{+\infty}
d\tau \exp{\left(-i\tau\log{A} \right)}\frac{\pi}{\cosh(\pi\tau)}\nn\\
& & = \frac{1}{\sqrt{A}}\frac{1}{2\pi}\frac{\pi}{\cosh\left(\frac{\log{A}}{2} \right)}= \frac{1}{\sqrt{A}}\frac{1}{2}\frac{1}{\frac{e^{\frac{1}{2}\log{A}}+e^{-\frac{1}{2}\log{A}}}{2}}\nn\\
& & = \frac{1}{\sqrt{A}}\frac{1}{\sqrt{A}+\frac{1}{\sqrt{A}}}=\frac{1}{1+A}\,,\quad {\rm c.q.d.} 
\eea}

\noi
Taking $N$-derivatives with respect to $A$ in this identity reproduces Eq.~\rf{eq:MBaidn}.

We shall next evaluate the Mellin transform of $\frac{1}{(1+A)^N}$  and show that
\be
\int_0^\infty dA\ A^{s-1} \frac{1}{(1+A)^N}=\frac{\Gamma(s)\Gamma(N-s)}{\Gamma(N)}\,.
\ee
We do that by applying Ramanujan's Master Theorem to the Taylor expansion:
\be
\frac{1}{(1+A)^N}=\sum_{k=0\,,1\,,2\dots} (-1)^k \left[\frac{\Gamma(N+k)}{\Gamma(N)\Gamma(k+1)}\right] A^{k}\,,
\ee
from which Ramanujan allows us to conclude  that

{\setl
\bea
\int_0^\infty dA\ A^{s-1} \frac{1}{(1+A)^N} & = & \Gamma(s)\Gamma(1-s)\times \left[\frac{\Gamma(N-s)}{\Gamma(N)\Gamma(-s+1)}\right]\\
 & = & \frac{\Gamma(s)\Gamma(N-s)}{\Gamma(N)}\,,\quad {\rm c.q.d.}\,.
\eea}
\vspace*{-0.5cm}
\section{\normalsize Positivity Properties of the Mellin Moments}
\setcounter{equation}{0}
\def\theequation{\Alph{section}.\arabic{equation}}

\noi
Because of the positivity property of the spectral function $\frac{1}{\pi}\Imm\Pi(t)$ the Mellin Moments $\cM(-N)$ which, here, for convenience, we write as follows
\be
\Sigma(N)=\int_{t_0}^\infty\frac{dt}{t_0}\left(\frac{t_0}{t} \right)^{2+N}\frac{1}{\pi}\Imm\Pi(t)\,,\quad N=0,1,2,\dots\,,
\ee
must satisfy certain constraints which we next discuss. Notice that with this definition:
\be
\cM(-n)\equiv \Sigma(N=n)\,.
\ee

It is useful to change variables slightly: set
\be
z=\frac{t_0}{t}\,,\quad \frac{dt}{t_0}=-\frac{dz}{z^2}\,,
\ee
and, therefore,
\be
 \Sigma(N)=\int_0^1 dz z^N \frac{1}{\pi}\Imm\Pi\left(\frac{1}{z}t_0\right)\,.
\ee
The positivity constraints follow from the fact that
\be
\sum_{N,N'}\left[\int_0^1 dz z^{N+N'} \frac{1}{\pi}\Imm\Pi\left(\frac{1}{z}t_0\right)\right]\xi_N \xi{_N'}\ge 0\,,
\ee
where $\xi_N$ and $\xi{_N'}$ are the components of arbitrary positive real vectors. This implies that the matrix 
\be
\Sigma(N,N')\equiv \int_0^1 dz z^{N+N'} \frac{1}{\pi}\Imm\Pi\left(\frac{1}{z}t_0\right)\,,
\ee
must be positive definite. The relevant constraints are then the following:

\begin{itemize}
	\item $N=N'=0$:
	
\be
	\Sigma(0)\ge 0\,.
\ee
	
	\item $(N,N')= 0,1$
	
\be
		\Sigma(0)\ge 0\,,\quad 	\Sigma(1)\ge 0\,,\quad
	\Sigma(1)\le \Sigma(0)\,.  
\ee

\item $(N,N')= 0,1,2$

\be
	\hspace*{-0.25cm}	\Sigma(0)\ge 0\,,\quad \Sigma(1)\ge 0\,,\quad \Sigma(2)\ge 0\,,\quad\Sigma(1)\le \Sigma(0)\,,\quad\Sigma(2)\le \Sigma(1)\,,
\quad
\Sigma(0)\Sigma(2)\ge [\Sigma(1)]^2\,.
\ee

\item $(N,N')= 0,1,2,3$

\be
		\Sigma(0)\ge 0\,,\quad \Sigma(1)\ge 0\,,\quad \Sigma(2)\ge 0\,,\quad\Sigma(3)\ge 0\,,
\ee
\be		\quad\Sigma(1)\le \Sigma(0)\,,\quad\Sigma(2)\le \Sigma(1)\,,\quad\Sigma(3)\le \Sigma(2)\,,
\ee
\be
\Sigma(0)\Sigma(2)\ge [\Sigma(1)]^2\,,\quad\Sigma(1)\Sigma(3)\ge [\Sigma(2)]^2\,,
\ee
and
\be
[\Sigma(0)-\Sigma(1)][\Sigma(2)-\Sigma(3)]\ge
[\Sigma(1)-\Sigma(2)]^2\,.
\ee

\end{itemize}

LQCD determinations of Mellin Moments should be consistent with  these constraints.

\end{appendix}

\vspace*{1.2cm}

\end{document}